\documentclass[aps,pre,twocolumn,amsmath,amsfonts,amssymb,floatfix,superscriptaddress,nofootinbib,notitlepage,10pt]{revtex4-1}
\usepackage{graphicx,amsmath,amssymb,amsfonts,bm, color}
\usepackage{todonotes}

\usepackage{xcolor,hyperref}
\hypersetup{
   colorlinks,
   linkcolor={blue!50!black},
   citecolor={blue!50!black},
   urlcolor={blue!80!black}
}

\newcommand{\beq}{\begin{equation}}
\newcommand{\eeq}{\end{equation}}
\newcommand{\beqnn}{\begin{equation*}}
\newcommand{\eeqnn}{\end{equation*}}
\newcommand{\bea}{\begin{eqnarray}}
\newcommand{\eea}{\end{eqnarray}}
\newcommand{\beann}{\begin{eqnarray*}}
\newcommand{\eeann}{\end{eqnarray*}}

\newcommand{\Hes}{\mathcal{M}}
\newcommand{\ev}{\vec{\psi}}
\newcommand{\dbar}{{\,\mathchar'26\mkern-12mu d}}

\begin{document}

\title{Statistical mechanics of local force dipole responses in computer glasses}
\author{Corrado Rainone}
\thanks{c.rainone@uva.nl}
\affiliation{Institute for Theoretical Physics, University of Amsterdam, Science Park 904, 1098 XH Amsterdam, The Netherlands}
\author{Eran Bouchbinder}
\affiliation{Chemical and Biological Physics Department, Weizmann Institute of Science, Rehovot 7610001, Israel}
\author{Edan Lerner}
\affiliation{Institute for Theoretical Physics, University of Amsterdam, Science Park 904, 1098 XH Amsterdam, The Netherlands}

\begin{abstract}
Soft quasilocalized modes (QLMs) are universally featured by structural glasses quenched from a melt, and are supposedly involved in a number of glassy anomalies such as the low temperature scaling of their thermal conductivity and specific heat, and sound attenuation at intermediate frequencies. In computer glasses, QLMs may assume the form of harmonic vibrational modes under a narrow set of circumstances, however direct access to their full distribution over frequency is hindered by hybridizations of QLMs with other low-frequency modes (e.g.~phonons). Previous studies to overcome this issue have demonstrated that the response of a glass to local force dipoles serves as a good proxy for its QLMs; we therefore study here the statistical-mechanical properties of the responses to local force dipoles in computer glasses, over a large range of glass stabilities and in various spatial dimensions, with the goal of revealing properties of the yet-inaccessible full distribution of QLMs' frequencies. We find that, as opposed to the spatial-dimension-independent universal distribution of QLMs' frequencies~$\omega$ (and, consequently, also of their stiffness $\kappa\!=\!\omega^2$), the distribution of stiffnesses associated with responses to local force dipoles features a (weak) dependence on spatial dimension. We rationalize this dependence by introducing a lattice model that incorporates both the real-space profiles of QLMs --- associated with dimension-dependent long-range elastic fields --- and the universal statistical properties of their frequencies. Finally, we discuss possible connections between our findings and basic aspects of the glass transition problem, and to finite-size effects in plastic activity of ultrastable glasses.
\end{abstract}

\maketitle

\section{Introduction}
\label{sec:intro}
\begin{figure}[!ht]
\includegraphics[width=0.5\textwidth]{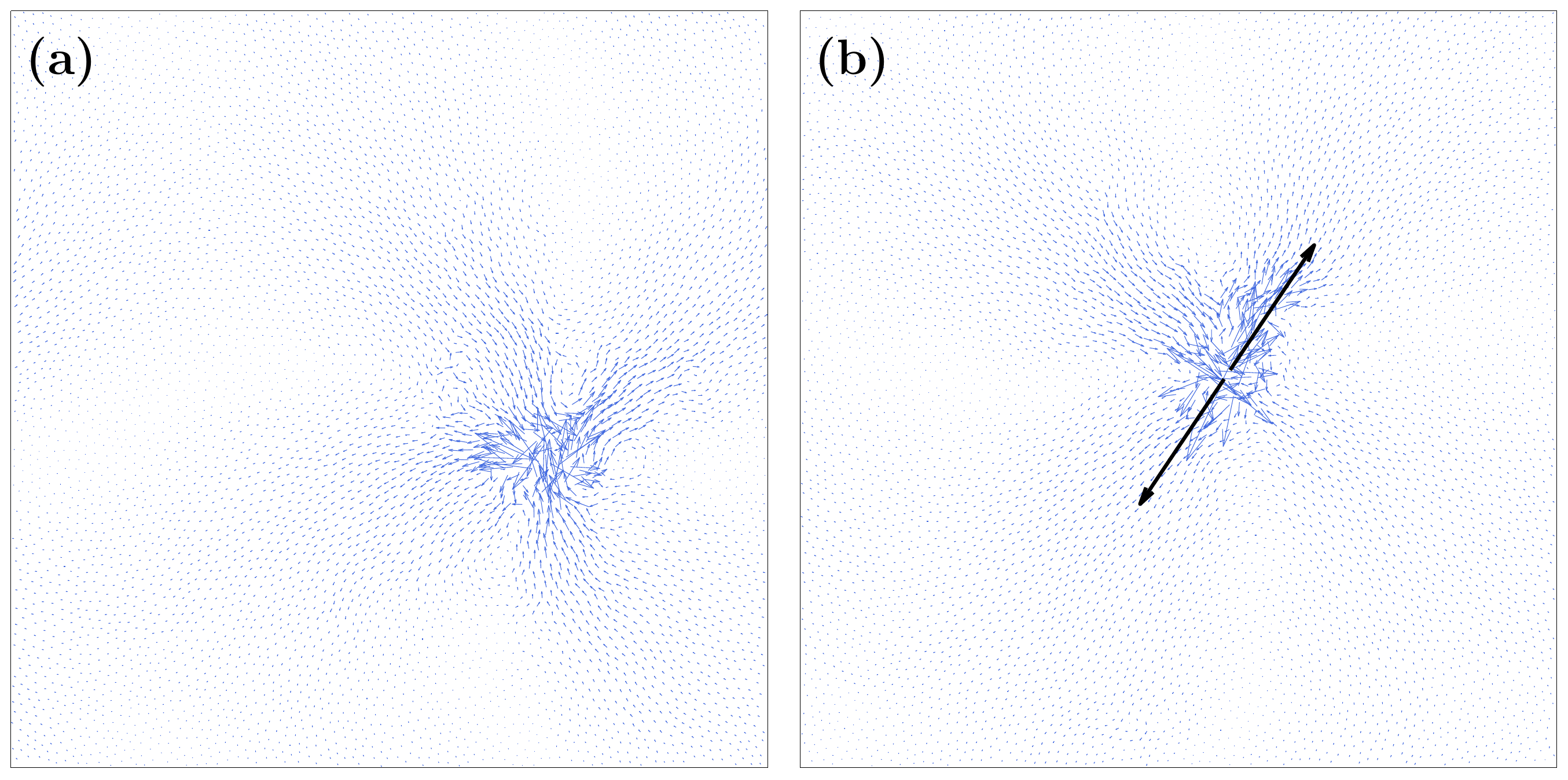}
\caption{\footnotesize (a) A low-frequency quasilocalized mode (QLM) observed in a 2D computer glass of $N\!=\!6400$ particles. (b) The normalized linear response to the local force dipole represented by the thick black arrows, and applied to the same computer glass model of panel (a) (but not at the same location of the core of the QLM shown in panel (a)). Both fields share common spatial structures, namely a disordered core decorated by algebraically-decaying elastic far fields of magnitude $\sim\!r^{1-\dbar}$ at distance $r$ away from the core, in $\dbar$ spatial dimensions. In this work we exploit the similarity between these two objects and explore the statistical mechanics of responses to local force dipoles. Details about the models and methods employed can be found in Section~\ref{sec:methods}.}
\label{fig1}
\end{figure}

All solids, whether crystalline or amorphous, can be described as elastic continua~\cite{59LL} at sufficiently large lengthscales. Accordingly, the low-frequency, long-wavelength harmonic vibrations are known to be phononic in nature for both of these classes of systems~\cite{AshcroftMermin}. In amorphous solids, however, another population of low-frequency excitations, which conversely are \emph{nonphononic} and quasilocalized in nature, exists~\cite{SL91,LS91}, see example in Fig.~\ref{fig1}a. Here and in what follows, we refer to these modes as quasilocalized modes (QLMs). QLMs are proposed to be the microscopic players behind a host of thermodynamic and kinetic anomalies (i.e.~experimental observations not explained by Debye theory) of low-temperature glasses~\cite{ZP71,AHV72,P72}, and have also been proposed to play a key role in equilibrium and nonequilibrium phenomena such as plastic deformation under shear or compressive strains~\cite{manningliu11,RSSL14,TMT10,KLP10b} and dynamical heterogeneities in the supercooled liquid melt~\cite{schober_1993_numerics,BW07,WPHR08,WPHR09,BW09} those glasses are prepared from.

The possibility to observe QLMs in the vibrational spectrum depends on them escaping hybridization with phonons \cite{BL18} or other QLMs \cite{kapteijns2019nonlinear}. Phononic excitations in finite-size computer glasses can be predicted and observed to be grouped into bands, each of them characterized by a frequency and a typical spectral width~\cite{BL18}. It is only below or in between these bands that nonphononic excitations can be realized as harmonic modes. When phononic and nonphononic vibrations share comparable frequencies $\omega$, they hybridize and the extended phononic background obscures the localized components: while localized soft structures are still embedded inside the glass, they become unobservable via harmonic spectral analyses~\cite{BL18}. Phonon bands get closer together when the system size is increased, and in the thermodynamic limit they are predicted to merge~\cite{BL18} and form the celebrated Debye density of states (DOS), $D_{\rm Debye}(\omega)\!\sim\!\omega^{\dbar-1}$ in $\dbar$ spatial dimensions. Creating a frequency window in computer glasses, wherein soft localized excitations can be observed as harmonic modes, and the low-frequency tail of their DOS ${\cal D}(\omega)\!\sim\!\omega^4$ subsequently calculated, amounts therefore to making the right choice of system and ensemble sizes~\cite{LDB16,KBL18,SMI18}. However, even when the universal tails of the QLM DOS are exposed by a harmonic analysis, the full form of the QLM DOS cannot be generally observed.

Assuming QLMs are relevant and important for glass physics, one should look for other possible ways to infer properties of their DOS from their impact on macroscopic observables. For example, the scaling form of the specific heat at low temperatures is indicative of the Debye DOS in crystalline materials~\cite{AshcroftMermin}. Similarly, in \cite{ZLBB17} it was shown that the form of the distribution of local, pairwise contributions to the specific heat in computer glasses indicates the universal $\omega^4$ low-frequency scaling of the DOS of QLMs (however, not the full QLM DOS). In this work, we explore a different approach, suggesting that the \emph{normalized} responses of a glass to local force dipoles --- referred to in what follows as \emph{dipole responses} --- are good proxies for QLMs, as demonstrated in Fig.~\ref{fig1} and discussed at length in \cite{LB18,RBL19}. Building on this similarity between dipole responses and QLMs, we study the statistical properties of the stiffnesses $\kappa$ associated with dipole responses, in an effort to shed light on the form of the full DOS of QLMs and its dramatic dependence on glass preparation protocol --- which, as explained above, are not accessible via harmonic analyses.

The suggestion that dipole responses are faithful representatives of QLMs was explored previously in \cite{LB18,RBL19}. In \cite{LB18} it was shown that QLMs and dipole responses, calculated in glasses made with the same preparation protocol, share the same system-size dependence (whose form depends, in turn, on spatial dimension), that stems from the algebraic tails of both quasilocalized objects. More recently, in \cite{RBL19} it was shown that the characteristic frequency scale of dipole responses is linked to their characteristic core size, and to the core size of QLMs. A similar proposition regarding the anomalous modes that emerge near the unjamming point \cite{ohern2003,liu_review,van_hecke_review} was put forward in \cite{new_variational_argument_epl_2016}. Here we go beyond these previous efforts, and study the entire distribution $p(\kappa)$ of the stiffnesses $\kappa$ (defined exactly below) associated with dipole responses. We show that $p(\kappa)$ generally features a scaling, power-law regime at low stiffnesses, and a rapid, exponential decay at high stiffnesses; the behavior of $p(\kappa)$ at intermediate stiffnesses is determined by glass stability. The exponent $\eta$ of the power-law regime $p(\kappa)\!\sim\!\kappa^\eta$ is not dimension-independent as for the QLM DOS, but it is independent of the form of inter-particle interaction and related to the universal QLM DOS scaling form ${\cal D}(\omega)\!\sim\!\omega^4$, and to the real-space long-range elastic fields of QLMs, via a mechanism which we describe in terms of a simple model of random variables on a lattice. We subsequently report the characteristic stiffness scale $\kappa_g\!\equiv\!\langle\kappa\rangle$ for a wide range of glass stabilities (wider than what was achieved in~\cite{RBL19}), and discuss the implications of this finding vis-\`a-vis the glass transition problem.

This paper is organized as follows; in section~\ref{sec:methods} we report the models and protocols used to prepare the glass samples; in~\ref{sec:statistics} we briefly re-introduce the dipole response and then report its statistics at various space dimensionalities, and introduce a simple, but illuminating, lattice model of it; in section~\ref{sec:stiffening} we calculate the bulk average of these dipole stiffnesses and report its dependence on glass stability down to the ultra-stable glass regime; section~\ref{sec:discussion} offers a discussion of our results and of future research perspectives. We conclude in section~\ref{sec:summary} with a brief summary of our main results.
\begin{figure*}[!ht]
\includegraphics[width=0.9\textwidth]{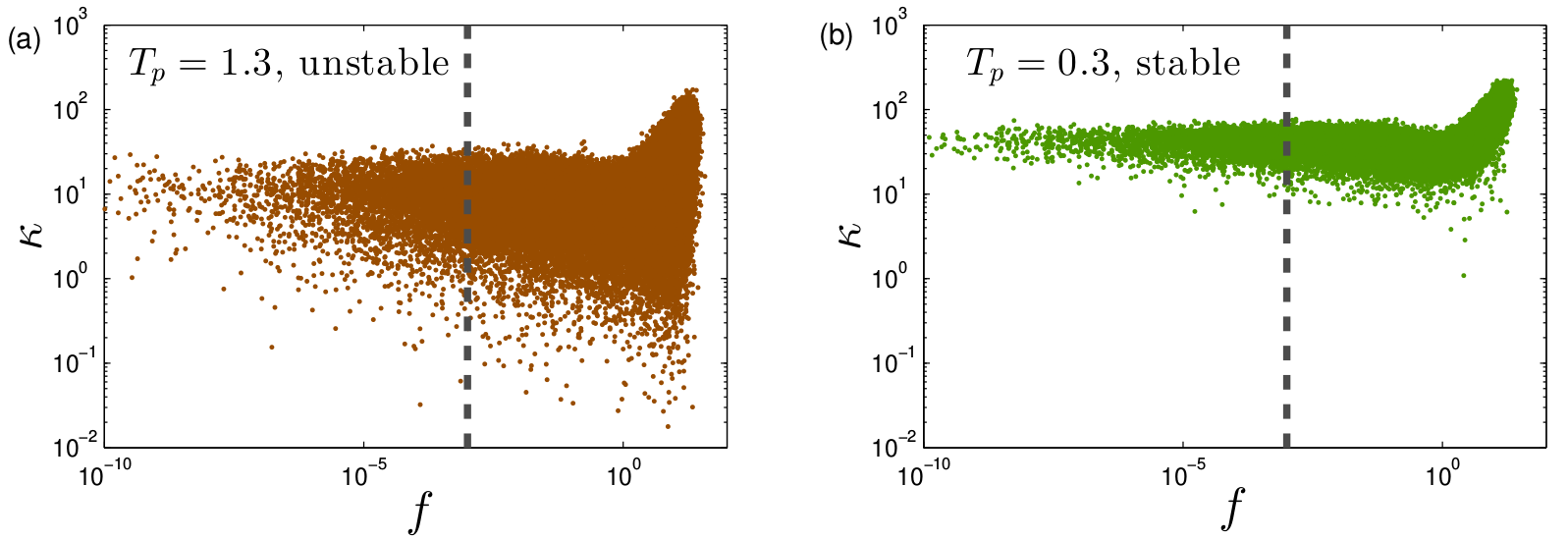}
\caption{\footnotesize Scatter plots of dipole stiffnesses $\kappa$ (defined in Eq.~(\ref{eq:ka}), where it is denoted as $\kappa_{ij}$) vs.~the pairwise forces $f$, measured for the POLY model with (a) $T_p\!=\!1.3$ and (b) $T_p\!=\!0.3$. $\kappa$ tends to increase for pairs that are strongly pressed together, but features $f$-independent statistics below a threshold of $10^{-3}$, marked by the vertical dashed lines.}
\label{filtering_fig}
\end{figure*}

\section{Models and methods}
\label{sec:methods}
We prepare athermal glassy configurations through numerical simulation of three computer glass forming models. The first model is a 50:50 binary mixture of `large' and `small' particles, interacting via a pairwise, inverse power-law potential $\sim\! r^{-10}$, with $r$ denoting the pairwise distance between particles. Details of this model can be found in \cite{KBL18,LB18}. Glassy samples of this model were created using two protocols: either by equilibrating states at a high parent temperature $T_p\!=\!1.0$ (in terms of the microscopic units of temperature detailed in \cite{LB18}), followed by a rapid quench by means of an energy minimization (see data of Fig.~\ref{fig:finitesize}), or by a continuous quench from the same parent temperature at a rate of $10^{-3}$ (see data of Fig.~\ref{fig:spatial_dependence_glass}). We used two, three, and four dimensional (2D, 3D, \& 4D) variants of this model, which is referred to in what follows as the IPL model.

The second computer glass forming model is a variant of the model proposed in~\cite{NBC17}: a system of continuously polydisperse disks with pairwise, inverse power-law interactions (same as for the IPL model), optimized for simulation with the Swap Monte Carlo~\cite{GrigeraParisi01,NBC17,GKPP15} (SMC) algorithm, which enables the preparation of athermal glassy configurations with a very wide range of mechanical and kinetic stabilities. We created glassy samples of this computer glass former in two and three dimensions (2D \& 3D). Thanks to the heightened efficiency of Swap Monte Carlo in 2D~\cite{BCN18,BCK19}, the degree of supercooling of our systems in 2D is significantly greater compared to our most deeply supercooled 3D systems: if $T_{\mbox{\tiny onset}}$ marks the onset parent temperature of the plateau of the athermal shear modulus (see, e.g., Refs.~\cite{RBL19,boring_paper}) of underlying inherent states of equilibrium configurations, then in 2D we equilibrate supercooled samples down to 15\% of $T_{\mbox{\tiny onset}}$, whereas in 3D we only reach 33\% of $T_{\mbox{\tiny onset}}$. Glassy samples were prepared by an energy minimization of configurations equilibrated at various parent temperatures $T_p$. We employed the filtering procedure presented in \cite{boring_paper} to handle sample-to-sample fluctuations induced by different realization of random particle diameters. We refer to this model as POLY.

The third computer glass forming model is a version the celebrated Kob-Andersen binary mixture~\cite{KobAndersen95}, which we henceforth refer to as KABLJ. Details about this model are spelled out in~\cite{boring_paper}. Glasses were preparaed by a fast quench (by means of energy minimization) from the equilibrium parent temperature $T_p\!=\!0.45$ (in terms of the units spelled out in~\cite{boring_paper}).

\section{Statistical mechanics of dipole responses}
\label{sec:statistics}

\subsection{Definitions and notations}
We launch the discussion with a brief recap of the definitions of dipole responses and their associates stiffness.

Given a system of $N$ particles whose energy is given by a sum of pairwise, radially-symmetric interactions, $U\! = \!\sum_{i<j}\varphi_{ij}(r_{ij})$, with $r_{ij}$ denoting a pairwise distance, a force dipole on the pair $i,j$ can be defined as
\begin{equation}
\vec{d}_{ij} \equiv \frac{\partial \varphi_{ij}}{\partial \vec{x}},
\end{equation}
where $\vec{x}$ is the $N\times \dbar$-dimension vector of all particle coordinates, and $\vec{d}_{ij}$ is a vector of the same dimension, but only features non-zero components on the $i,j$ pair of particles. The linear response $\vec{z}_{ij}$ of the system to such a force is given by
\begin{equation}
\vec{z}_{ij} = \mathcal{M}^{-1}\cdot\vec{d}_{ij},
\end{equation}
where $\mathcal{M}$ is the matrix of second derivatives (Hessian) of the potential energy, $\mathcal{M}\equiv \frac{\partial U}{\partial \vec{x}\partial \vec{x}}$, and the dot ($\cdot$) denotes a single contraction over particle indices and spatial dimensions. An example of a dipole response in a 2D computer glass is presented in Fig.~\ref{fig1}b.

A \emph{dipole stiffness} $\kappa_{ij}$ can be associated with every dipole response $\vec{z}_{ij}$ as
\begin{equation}
\kappa_{ij} = \frac{\vec{z}_{ij}\cdot\Hes\cdot\vec{z}_{ij}}{\vec{z}_{ij}\cdot\vec{z}_{ij}}=\frac{\vec{d}_{ij}\cdot \Hes^{-1}\cdot\vec{d}_{ij}}{\vec{d}_{ij}\cdot\Hes^{-2}\cdot\vec{d}_{ij}}\,.
\label{eq:ka}
\end{equation}
In the models studied, the dipole stiffnesses $\kappa_{ij}$ can depend on the pairwise force $f_{ij}\!\equiv\!-\partial\varphi_{ij}/\partial r_{ij}$ exerted between the pair of particles $i,j$; in particular, as also shown in \cite{RBL19}, we find that applying a force dipole $\vec{d}_{ij}$ on pairs with large pairwise forces results in responses with large associated stiffnesses (see Fig.~\ref{filtering_fig}), that are presumably not good proxies of \emph{soft} quasilocalized modes. However, for pairs with forces below a chosen threshold of $10^{-3}$, marked as the vertical dashed lines in Fig.~\ref{filtering_fig}, the statistics of the associated dipole response stiffnesses become independent of the force. For this reason, in all the following analyses we only consider the dipole responses of pairs $i,j$ for which the pairwise force is \emph{smaller} than the threshold (with the exception of the KABLJ model, wherein the attractive part of the potential can result in \emph{negative} pair forces, making such a filtering procedure meaningless).

We next turn to studying the statistical properties of the dipole stiffnesses $\kappa$.
\begin{figure*}
 \includegraphics[width=1.0\textwidth]{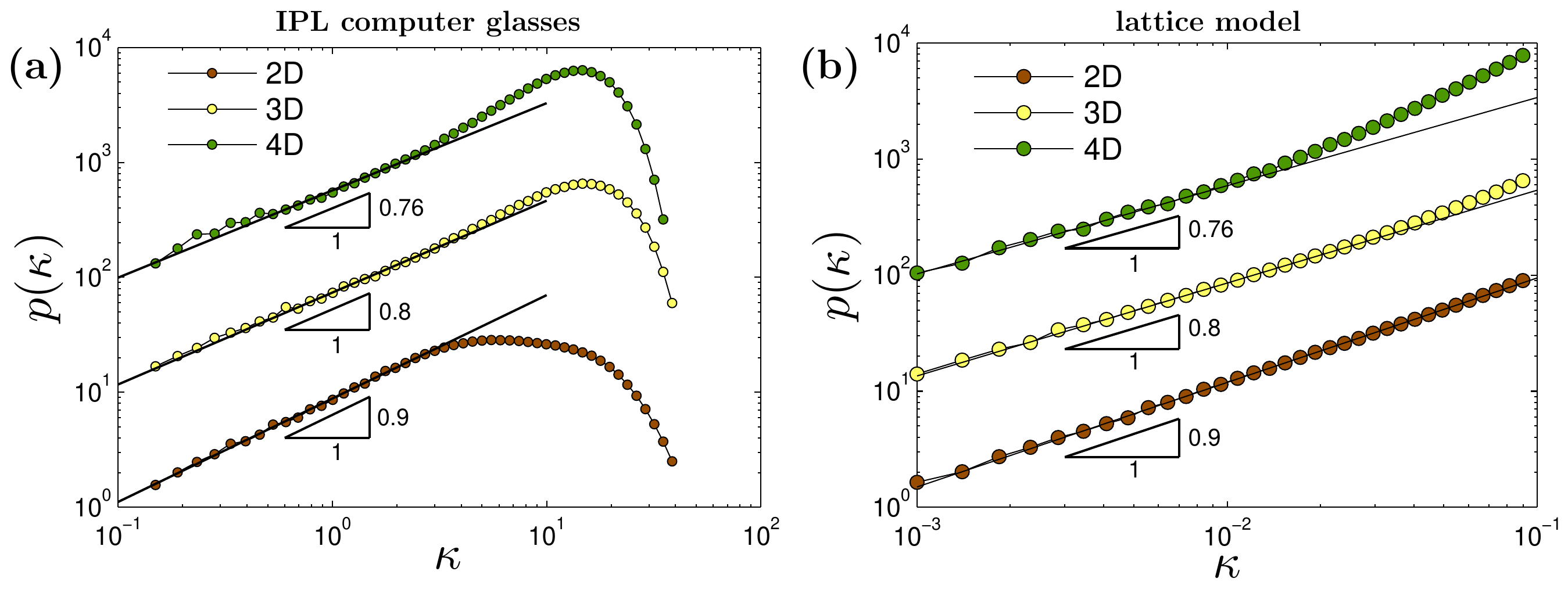}
 \caption{(a) The distributions $p(\kappa)$ of dipole stiffnesses for 2D ($N\!=\!6400$, brown), 3D ($N\!=\!128000$, yellow), and 4D ($N\!=\!40000$, green) glassy solids of the IPL model. The solid black lines have slopes as indicated in the figure, which correspond to the predictions of the lattice model, whose results are presented in panel (b) and which is discussed in Sect.~\ref{subsec:model}. (b) $p(\kappa)$ as it results from simulation of the lattice model discussed in the text, for dimensions $\dbar=2,3$ and $4$.}
 \label{fig:spatial_dependence_glass}
\end{figure*}

\subsection{Spatial-dimension dependence of $\kappa$ statistics}

We first present in Fig.~\ref{fig:spatial_dependence_glass}a the distributions $p(\kappa)$ for the IPL model, measured in the ensembles of glasses quenched continuously at a finite rate (see details in Sect.~\ref{sec:methods}), and for different spatial dimensions. Note that in order to obtain converged distributions $p(\kappa)$, large glasses must be employed (see caption of Fig.~\ref{fig:spatial_dependence_glass}). The finite-size effects affecting these measurements are discussed in Appendix~\ref{app:finitesize}.

Naively, a simple transformation from QLM frequencies $\omega$ to stiffnesses in the form $\kappa\!=\!\omega^2$ would suggest $p(\kappa)\!\sim\!\kappa^{3/2}$, independently of spatial dimension, based on the universal QLM DOS ${\cal D}(\omega)\!\sim\!\omega^4$~\cite{KBL18}, and assuming each force dipole couples to a \emph{single} QLM. We find instead that at small stiffnesses $p(\kappa)\!\sim\!\kappa^\eta$ with a dimension-dependent exponent $\eta$ ranging between $\approx\!0.90$ in 2D, to $\approx\!0.76$ in 4D. The values of the dimension-dependent exponent $\eta$ in the power-law fits in Fig.~\ref{fig:spatial_dependence_glass}a are not a result of looking for the best fits, but rather correspond to the predictions of a simple lattice model, the results of which are shown in Fig.~\ref{fig:spatial_dependence_glass}b, which we discuss next.

\subsection{Lattice model}
\label{subsec:model}

We aim at constructing a simple model that will allow us to gain physical insight into the observed dipole stiffnesses statistics, and, in particular, its observed $\dbar$ dependence. We start by defining
\begin{equation}
\kappa = \frac{\vec{d}\cdot \Hes^{-1}\cdot\vec{d}}{\vec{d}\cdot\Hes^{-2}\cdot\vec{d}} \,\equiv\, \frac{a}{b}\,,
\end{equation}
where we suppressed the indices $i,j$ for notational simplicity, while keeping in mind that $\kappa,a,b$ and $\vec{d}$ are all quantities associated with some pair $i,j$ of interacting particles.Using Eq.~\eqref{eq:ka} and the spectral decomposition of the Hessian matrix ${\cal M}\!=\!\sum_\ell \omega_\ell^{2}\vec{\psi}_\ell\otimes\vec{\psi}_\ell$, one obtains
\beq
a = \sum_{\ell} \frac{(\vec{d}\cdot\vec{\psi}_\ell)^2}{\omega_\ell^2},\qquad b = \sum_{\ell} \frac{(\vec{d}\cdot\vec{\psi}_\ell)^2}{\omega_\ell^4}\ ,
\label{eq:ab_spectral}
\eeq
where the sums run over all eigenmodes $\vec{\psi}_\ell$ of ${\cal M}$.

We next assume that the sums can be restricted to the eigenvectors associated with QLMs because force dipoles $\vec{d}$ have a negligible scalar product with phononic excitations (we recall that the dipole response is indeed explicitly defined as such because of its real-space resemblance to QLMs~\cite{LB18}). With this assumption in mind, one can treat the frequencies $\omega_\ell$ as random variables extracted from the ${\cal D}(\omega)\!\sim\!\omega^4$ distribution. Next, we estimate the scalar products $\ev\cdot\vec{d}$ using the well-established spatial structure of QLMs. As the dipole vector $\vec{d}$ has the geometry of two forces pulling in opposite directions, it can be approximated as a gradient operator acting on the eigenmode $\ev$. As reported in~\cite{KBL18,RBL19}, $\ev$ is characterized by a disordered and localized core of size $\xi_{\mbox{\tiny QLM}}$, decorated with a long-range elastic power-law decay at a large distances $r$ away from it, whose form is
\begin{equation}
|\vec{\psi}|(r) \sim r^{-(\dbar-1)} \ .
\label{eq:farfield}
\end{equation}
Consequently, we have
\beq
(\vec{\psi}\cdot\vec{d})^2 \sim (\partial_r r^{-(\dbar-1)})^2 \sim r^{-2\dbar}\,,
\label{eq:ff_propagator}
\eeq
where $r$ is the Euclidean distance in $\dbar$ dimensions between the core of the quasilocalized mode $\vec{\psi}$ and the point at which the dipole force $\vec{d}$ is applied.

Taken together, we define a model of interacting random variables $\omega_i$ on a $\dbar$-dimensional cubic lattice that follow an independent and identical quartic power-law distribution
$$
p(\omega) = 5\omega^4,\ \omega\in[0,1],
$$
For each site $i$, we then define the quantities
\begin{eqnarray}
a_i &\equiv\ &\frac{1}{\omega^2_i} + \sum_{j\neq i}^N \frac{r_{ij}^{-2\dbar}}{\omega_j^2}\,, \label{eq:a_model}\\
b_i &\equiv\ &\frac{1}{\omega^4_i} + \sum_{j\neq i}^N \frac{r_{ij}^{-2\dbar}}{\omega_j^4}\,, \label{eq:b_model}\\
\kappa_i &\equiv &\frac{a_i}{b_i}\,,
\end{eqnarray}
where $r_{ij}$ is now the distance between the lattice sites $i$ and $j$, and $N$ is the total number of sites in the lattice. These definitions are meant to mimic the sums found in Eq.~\eqref{eq:ab_spectral}, and we have assumed that $\vec{\psi}_j\cdot\vec{d}$ equals unity when the dipole and the QLM's core are located at the lattice site.

We are interested in the probability distribution function of the $\kappa_i$s, and in particular in the exponent $\eta$ of its power-law regime, $p(\kappa) \simeq \kappa^\eta$ when $\kappa\to0$. We obtain it by simulating the model on a $\dbar$-cubic lattice in 2D (lattice side length $L=101$), 3D ($L=21$) and 4D ($L=11$). In order to suppress boundary effects, we choose the site $i$ to be at the very center of the lattice and compute $a_i$, $b_i$ and $\kappa_i$ according to the definitions above (i.e.~each realization of the model gives rise to a single value of $\kappa_i$). In order to obtain statistical convergence, we repeat this process for $N_{\mbox{\tiny samples}}\!=\!10^7$ lattices. The results are reported on in Fig.~\ref{fig:spatial_dependence_glass}b. It is observed that the power-law exponents $\eta(\dbar)$ in the model are consistent with those measured in computer glasses, cf.~panel (a), suggesting that the model properly captures the underlying physics giving rise to the $\dbar$ dependence of the exponent $\eta$.

We continue with some comments on our lattice model. As always, an exact solution of it would be desirable, in particular to obtain the function $\eta(\dbar)$ characterizing the spatial-dimension dependence of $\eta$. However, even in the absence of an exact solution for $\eta(\dbar)$, some understanding of the $\dbar$-dependence of the model can be achieved using simple statistical arguments, spelled out next.

An illuminating way of rewriting Eq.~\eqref{eq:a_model} is
\begin{equation}
a_i = \frac{1}{\omega^2_i} + \sum_{\mbox{\scriptsize shells}} r_{\mbox{\scriptsize shell}}^{-2\dbar} \sum_{j \in \textrm{shell}} \frac{1}{\omega_j^2} \equiv \frac{1}{\omega^2_i} + \sum_{\mbox{\scriptsize shells}} A_r \,,
\end{equation}
where we have split the sum over all sites into a sum over spherical shells of radius $r_{\mbox{\scriptsize shell}}$, and a sum over sites within those shells, which results in the definition $A_r\!\equiv\!r^{-2\dbar}\sum_{j\in\textrm{shell}}\frac{1}{\omega^2_j}$. A similar resummation can be written for Eq.~(\ref{eq:b_model}), namely
\begin{equation}\label{foo02}
b_i = \frac{1}{\omega^4_i} + \sum_{\mbox{\scriptsize shells}} r_{\mbox{\scriptsize shell}}^{-2\dbar} \sum_{j \in \textrm{shell}} \frac{1}{\omega_j^4} \equiv \frac{1}{\omega_i^4} + \sum_{\mbox{\scriptsize shells}} B_r \,,
\end{equation}
with $B_r\!\equiv\!r^{-2\dbar}\sum_{j\in\textrm{shell}}\frac{1}{\omega_j^4}$. 

We note next that the number of independent variables summed over in $A_r$ and in $B_r$ grows with their shells' radius as $\sim\! r^{\dbar-1}$. Thus, for large $r$ one expects the distribution $p(A_r)$ to converge to a Gaussian due to the law of large numbers, whose mean is expected to scale as $r^{-(\dbar+1)}$ due to the $\sim\! r^{-2\dbar}$ spatial decay of QLMs, combined with the $\sim\! r^{\dbar-1}$ scaling of the number of variables in a shell of radius $r$. 

The situation for $B_r$ is drastically different, since it consists of a sum over variables $\omega_i^{-4}$, which, together with the universal form ${\cal D}(\omega)\!\sim\!\omega^4$ from which the variables $\omega_i$ are drawn, leads to Levi Stable law for $p(B_r)$ that features a power-law tail for large $B_r$, and a characteristic scale $\sim\! r^{-(\dbar+1)}$ \cite{BOUCHAUD1990127}, namely
\begin{equation}\label{foo00}
p(B_r) \sim \left( \frac{B_r}{r^{-(\dbar+1)}}\right)^{-9/4}\,.
\end{equation}

As the spatial dimension $\dbar$ is increased, only the first shell of neighbors needs to be considered. We denote by $n_\dbar$ the number of members in the first shell; following the law of large numbers, $a_i$ will abide by a Gaussian distribution of mean $\sim\! n_\dbar$ and width $\sim\! \sqrt{n_\dbar}$. At the same time, $b_i$ will be distributed according to a Levi Stable law with mean $\sim\! n_\dbar$, but with a width $\sim\! n_\dbar^{4/5}\!\gg\!\sqrt{n_\dbar}$~\cite{BOUCHAUD1990127}, i.e.~much larger than the width of the (Gaussian) distribution of $a_i$. For these reasons, the stochastic nature of $a_i$ can be neglected, and then one expects
\begin{equation}\label{foo01}
\kappa_i \sim 1/b_i\,, \quad \mbox{for}\  \dbar \to \infty\,.
\end{equation}
Following Eqs.~(\ref{foo02}) and (\ref{foo00}), we expect $p(b)$ to inherent the Levi Stable law $\sim\! b^{-9/4}$. Therefore, together with Eq.~(\ref{foo01}), we expect that
\begin{equation}
p(\kappa) \sim \kappa^{1/4}\,, \quad \mbox{for}\ \dbar \to \infty\,,
\end{equation}
i.e.~substantially lower than any of the exponents we observed in our numerics. These arguments explain the downward trend reported in Fig.~\ref{fig:spatial_dependence_glass}b for the scaling exponent $\eta(\dbar)$ with increasing spatial dimension, and highlight the role of correlations between $a$ and $b$ in forming the observed distributions $p(\kappa)$ for their ratio $\kappa$. 

\begin{figure}[!ht]
\includegraphics[width=0.5\textwidth]{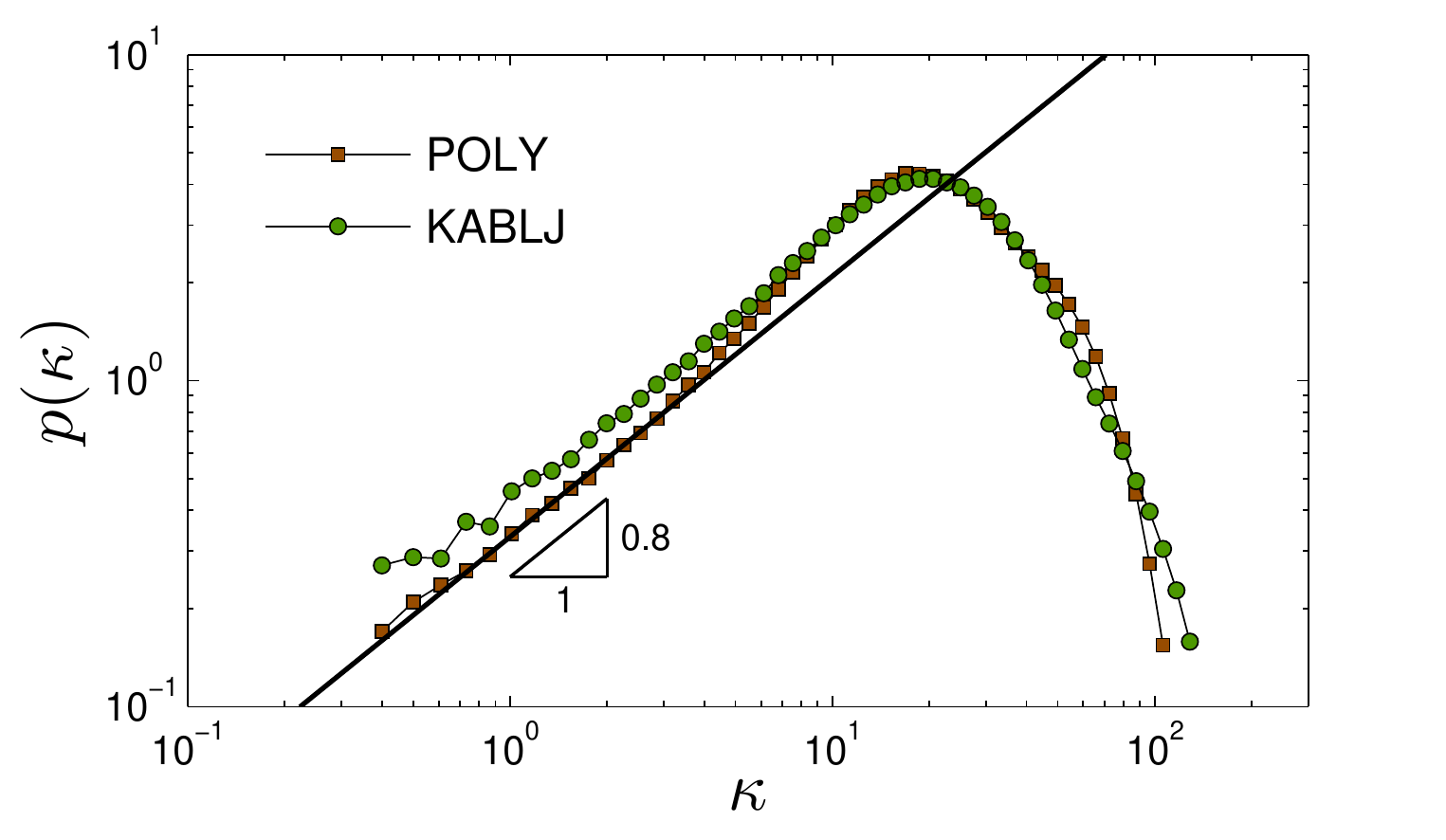}
\caption{\footnotesize The distribution of dipole stiffnesses $p(\kappa)$ measured in the POLY system with $N\!=\!2000$ and $T_p\!=\!0.50$, and the KABLJ system with $N\!=\!2000$ and $T_p\!=\!0.45$. We observe no dependence of the exponent on the nature of the microscopic interactions.}
\label{model_independence}
\end{figure}

\subsection{Glass-former independence}

The success of the lattice model to capture the physics behind the spatial-dimension-dependent scaling exponents $\eta(\dbar)$ suggests that the exponent is independent of the details of microscopic interactions. To test this prediction, we also measured $p(\kappa)$ in the KABLJ glass forming model (see Sect.~\ref{sec:methods} for details) and in the POLY system, and plot the results in Fig.~\ref{model_independence}. We indeed find the same exponent $\approx\!0.80$, demonstrating glass-former independence. Moreover, the results show the attractive nature of the pairwise interactions of the KABLJ model does not alter the scaling exponent. We note that the data shown for both models in Fig.~\ref{model_independence} were obtained by exciting random pairs of interacting particles, regardless of their pairwise forces.

\section{Protocol dependence of the dipole response statistics}
\label{sec:stiffening}

\begin{figure}[!ht]
 \includegraphics[width=0.5\textwidth]{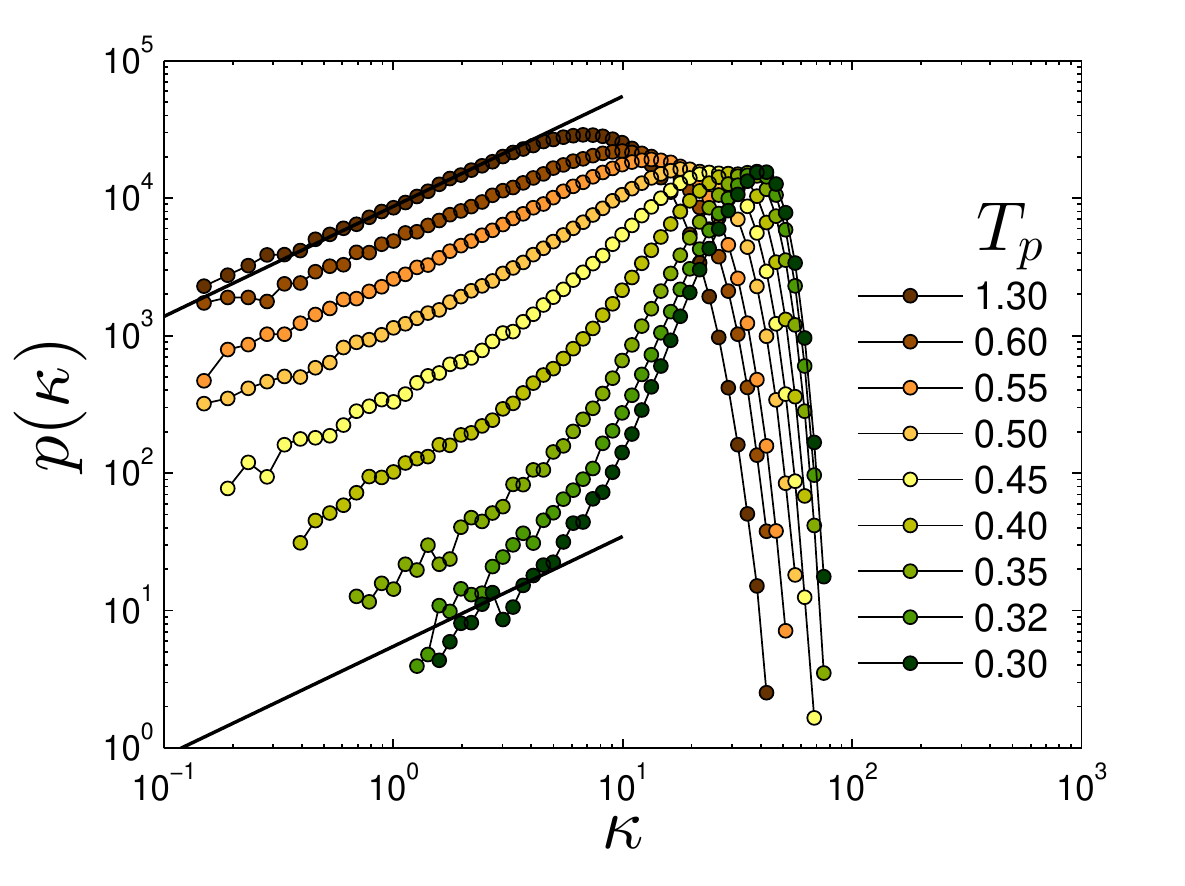}
 \caption{The distributions of dipole stiffnesses $p(\kappa)$ measured in the POLY system in 3D, employing systems of $N\!=\!16000$ for $T_p\!>0.55$, and systems of $N\!=\!2000$ for $T_p\!\le\!0.55$, for a broad range of $T_p$ as indicated in the legend. The black continuous lines represent the scaling $\sim\!\kappa^{0.80}$ as found for our lattice model in 3D (see Fig.~\ref{fig:spatial_dependence_glass}b).}
\label{fig:pdfk_Tp}
\end{figure}

After discussing the dependence of the scaling regime of $p(\kappa)$ on the spatial dimension, and establishing its independence of interaction details, we next focus on how $p(\kappa)$ depends on the \emph{protocol} employed to prepare the glass, i.e., in the case of the POLY system, on the equilibrium parent temperature $T_p$ from which the glassy configurations were instantaneously quenched. In Fig.~\ref{fig:pdfk_Tp}, we report $p(\kappa)$ for the POLY system, over a very large range of parent temperatures $T_p$. We find that the exponent characterizing the power-law regime appears to be roughly independent of $T_p$, similarly to what is observed in the QLM DOS, which features a $T_p$-independent scaling regime $\sim\!\omega^4$~\cite{RBL19,WNGBSF18}. In addition, the behavior of the \emph{prefactor} $A_\kappa(T_p)$ of the $p(\kappa)\!=\! A_\kappa\kappa^\eta$ scaling regime changes dramatically with $T_p$, by a factor of more than $10^3$. This feature is also akin to the behavior of the prefactor $A_g(T_p)$ of the QLM DOS ${\cal D}(\omega)=A_g\omega^4$, shown very recently \cite{RBL19} to vary by a similar order of magnitude over the same parent temperature range, for the same systems. These commonalities between $p(\kappa)$ and ${\cal D}(\omega)$ strengthen our suggestion that studying one can shed light on the other.

A third noticeable, and important, dependence of $p(\kappa)$ on the parent temperature $T_p$ is the upwards shift in its peak location with lowering $T_p$. This feature, as well as the stiffness scale that marks the end of the $p(\kappa)\!\sim\!\kappa^\eta$ scaling regime, are generally unobservable in ${\cal D}(\omega)$ obtained by harmonic analyses~\cite{footnote}, particularly in stable glasses, due to hybridizations of QLMs with phonons and other low-energy excitations \cite{LB18,kapteijns2019nonlinear}. We note that in~\cite{RBL19}, it has been proposed that the stiffness scale that marks the end of the $p(\kappa)\!\sim\!\kappa^\eta$ scaling regime features the same $T_p$ dependence as $\kappa_g$ (the mean of $p(\kappa)$, defined earlier).

The shift of the distributions $p(\kappa)$ to higher $\kappa$ with decreasing $T_p$ indicates the stiffening of the \emph{characteristic scale} of dipole stiffnesses, and is presumably also featured by the underlying full distribution ${\cal D}(\omega)$ of QLMs' frequencies. Indeed, in \cite{RBL19} information about the characteristic frequency scale of dipole responses $\omega_g\!\sim\!\sqrt{\kappa_g}$ was used to disentangle the stiffening of QLMs frequencies with decreasing $T_p$, from their annealing-induced depletion. It was further shown in \cite{RBL19} that $\omega_g$ is related to the linear size $\xi_{\mbox{\tiny QLM}}$ of QLMs' core, strengthening the connection between dipole responses and soft QLMs.

\begin{figure}[!ht]
\includegraphics[width=0.48\textwidth]{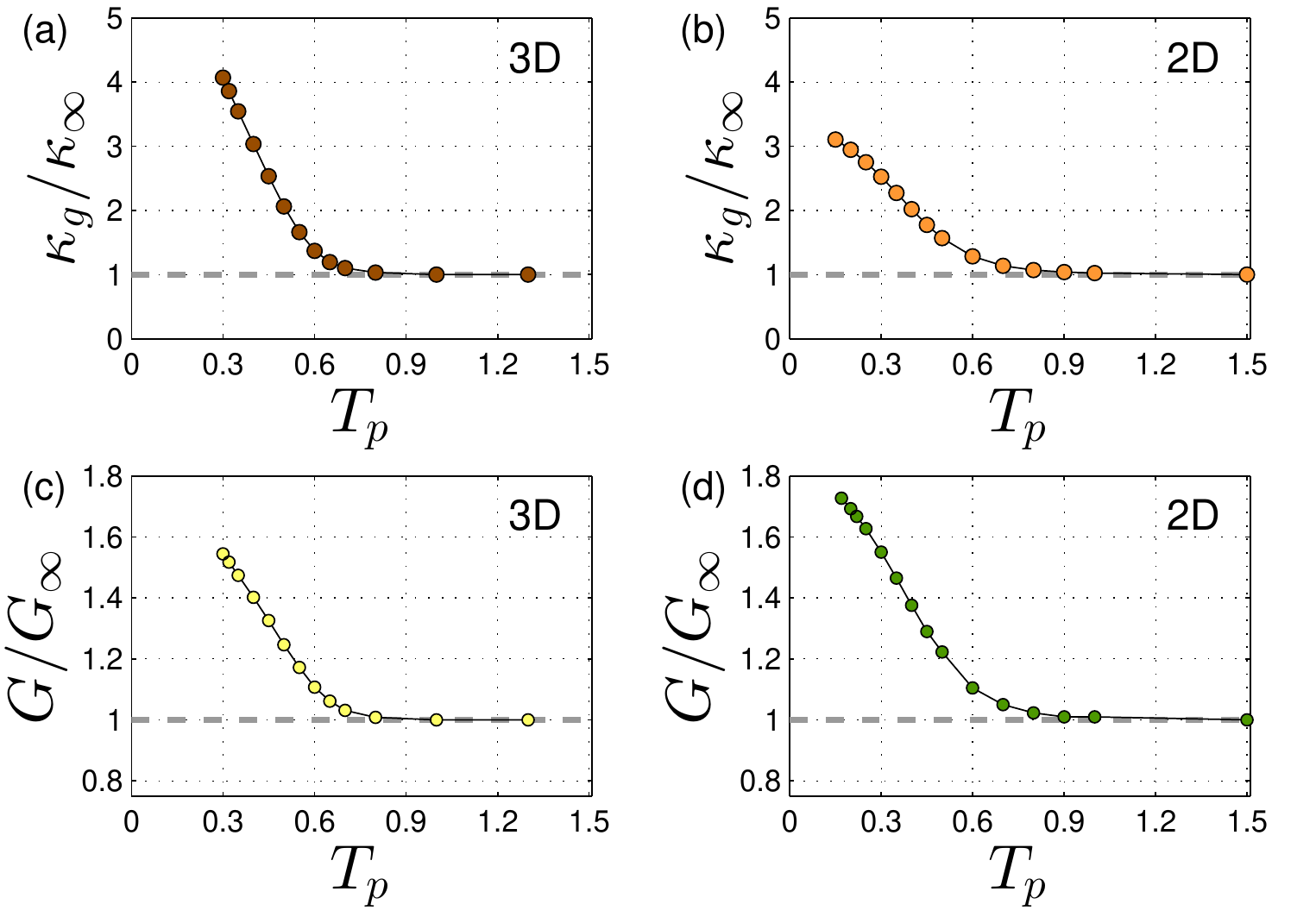}
\caption{Mean dipole stiffness $\kappa_g$, rescaled by its high-parent-temperature value $\kappa_\infty\!\equiv\!\kappa_g(T_p\!\to\!\infty)$ vs.~parent temperature $T_p$ for (a) 3D and (b) 2D systems of the POLY model. We also show the sample-to-sample mean athermal shear modulus $G$, rescaled by its high-parent-temperature value $G_\infty\!\equiv\! G(T_p\!\to\!\infty)$, and plotted against $T_p$ for (c) 3D and (d) 2D POLY systems. Notice the differences between the scales of the y-axes in panels (a)\&(b) vs.~panels (c)\&(d).}
\label{kappa_g_vs_Tp}
\end{figure}

In Fig.~\ref{kappa_g_vs_Tp} we show the averages $\kappa_g$ of the dipole stiffnesses, both for our 3D (panel (a)) and our 2D (panel (b)) glassy samples of the POLY model. Two main observations can be made; first, we can see that the curve in Fig.~\ref{kappa_g_vs_Tp}b seems to bend downwards at the lowest parent temperatures, an effect which is not visible in 3D, and can be attributed to the higher degree of supercooling we attained in 2D thanks to the higher efficiency of Swap Monte Carlo at that dimensionality~\cite{BCN18,BCK19}. This alludes to the possibility that the typical stiffness of QLMs do not grow indefinitely on supercooling, but rather saturate to a finite value, a phenomenon which could have wide-ranging implications for the glass transition problem, discussed in the next Section. Secondly, the increase of $\kappa_g$ is much more pronounced compared to the one found in other typical elastic stiffnesses of the system. We support this statement by reporting, again against $T_p$, the sample-to-sample athermal shear modulus $G$, again for both 2D and 3D, calculated using the atomistic framework of~\cite{LM06} and averaged over our entire ensembles of glassy samples. It is apparent that the effect of annealing on $\kappa_g$ is indeed much stronger, compared to the one it has on the macroscopic shear modulus: $\kappa_g$ stiffens by a factor of 3 in 2D, and a factor of more than 4 in 3D. Interestingly, this implies that the effects of annealing on glass physics are much more visible at the level of micro- and meso-scale elasticity (as captured by the dipole responses) than at the macroscopic level.
\begin{figure}[!ht]
\includegraphics[width=0.48\textwidth]{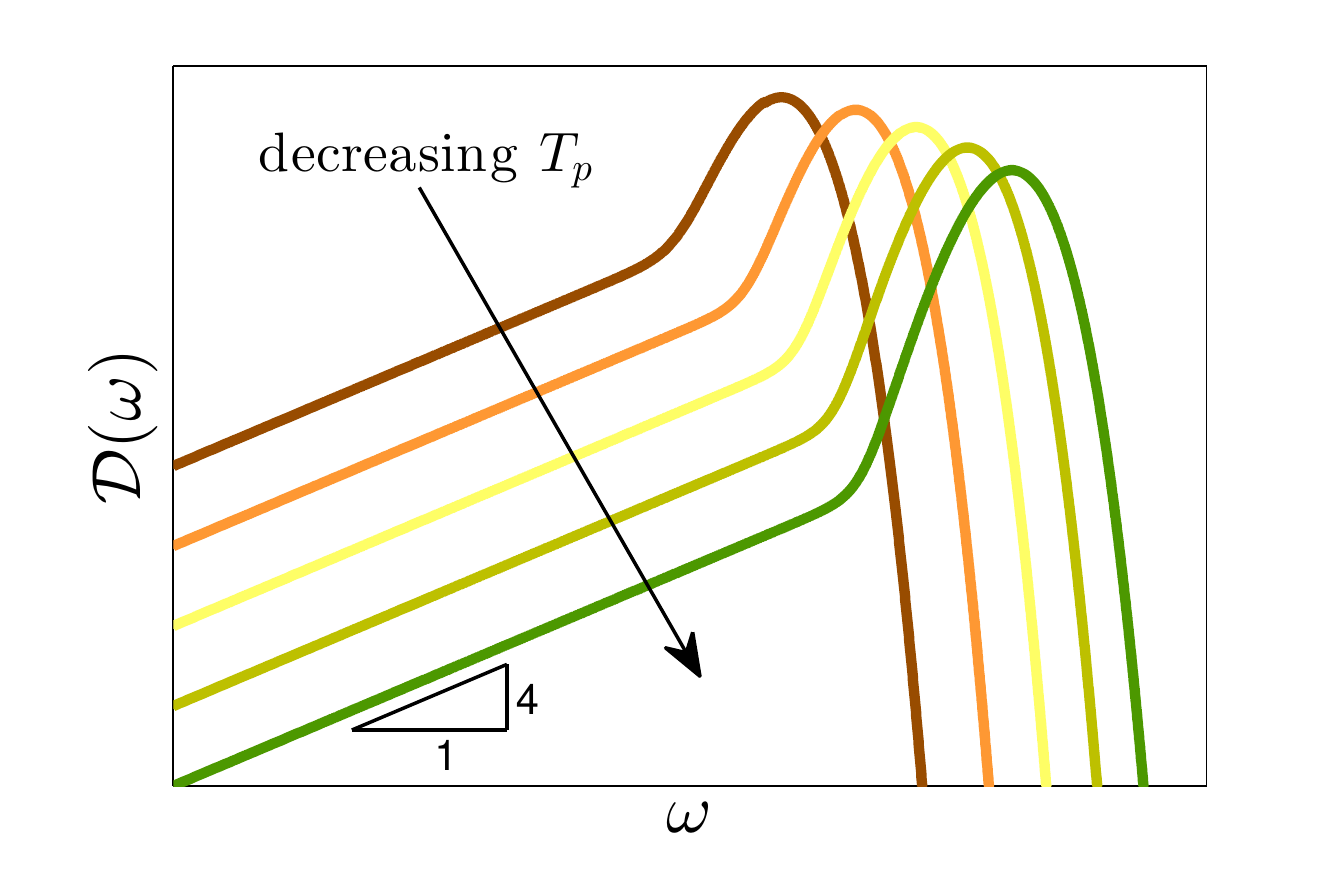}
\caption{\footnotesize A cartoon of the speculated form of the full distribution of soft, non-phononic excitations in glasses (QLMs), drawn based on the insights gained throughout this work.}
\label{fig:DOScartoon}
\end{figure}

Putting together all of the results collected so far, we can sketch a cartoon of how the full QLM DOS ${\cal D}(\omega)$ will change upon variations of $T_p$, for low $T_p$ values (presumably representative of laboratory glasses). It is already well known that the change in the scaling ($\omega\!\to\!0$) regime --- which takes the form ${\cal D}(\omega)\!=\! A_g\omega^4$ --- amounts to a reduction of its prefactor $A_g$~\cite{WNGBSF18,RBL19}, which is caused by both a depletion and a stiffening of its modes \cite{LB18,RBL19}. The frequency scale $\omega_g\!\sim\!\sqrt{\kappa_g}$ presumably separates the scaling regime from a peak regime, which we understand as corresponding to the peak regime of the distribution of dipole stiffnesses. These speculated features are represented in the cartoon shown in Fig.~\ref{fig:DOScartoon}. We stress that while the scaling regime of the QLM DOS has been observed in multiple works~\cite{LDB16,KBL18,WNGBSF18,RBL19}, this is not the case for the peak regime, due to the stronger pollution by hybridizations with phonons at those frequencies. At the moment, the only way to observe it, albeit indirectly, is through the dipole response statistics presented in this work.

\section{Discussion}
\label{sec:discussion}

The different ways in which the statistics of the QLM DOS respond to changes in $T_p$ is of interest concerning the longstanding riddle of the glass transition. Liquids are observed to become exceedingly viscous upon supercooling below their melting point~\cite{CavagnaLiq}. In the case of fragile glass formers, their viscosity $\mu$ dependence on temperature $T$ exceeds the naive prediction based on Arrhenius' law, $\mu\!\sim\!\exp(\Delta E/T)$, where $\Delta E$ represents a temperature-\emph{independent} activation energy. The steeper-than-Arrhenius temperature dependence of viscosity in fragile glass formers implies that a temperature-dependent activation energy $\Delta E(T)$, that increases upon cooling, must emerge \cite{tarjus_2004}. Based on the aforementioned observation that the most mobile regions of a supercooled liquid strongly correlate with the ones that feature quasilocalized soft modes~\cite{schober_1993_numerics,BW07,WPHR08,WPHR09,BW09}, it appears reasonable to assume that QLMs play a part in driving flow and viscosity in supercooled melts.

Accordingly, it was already suggested by some of us in~\cite{LB18}, that the characteristic scale of dipole stiffnesses $\kappa_g(T_p)$ controls the activation energy $\Delta E(T)$. In the vast landscape of theories of the glass transition, this idea can be located close to the one formulated by Dyre that the activation energy be essentially \emph{elastic} in nature~\cite{Dy06}, with the key difference that while Dyre refers to macroscopic elasticity as quantified by the high-frequency (`plateau') shear modulus as the observable that controls activation energies \cite{PhysRevB.53.2171,Dy06}, here we focus on mesoscale elasticity as captured by the characteristic stiffness of QLMs, argued in \cite{LB18,RBL19} to be given by $\kappa_g$. In this sense, the proposal of~\cite{LB18} is also close in spirit to the picture of Brito and Wyart, who envision the glass transition as being caused by the progressive stabilization of soft excitations found close to the jamming point of amorphous packings~\cite{BW07,BW09}, though in that case those excitations are collective, rather than quasilocalized.

Firmly establishing the connection between mesoscale elasticity in underlying inherent states --- as reflected by $\kappa_g$ --- and activation energies in the supercooled melt, is an important direction for future research. Assuming that such a connection indeed holds, we make two comments about its implications based on the data presented in Fig.~\ref{kappa_g_vs_Tp}; firstly, it appears clear that macroscopic elasticity cannot explain fragility due to its mild variation with temperature: in our 3D systems, we find that the shear modulus varies by less than 60\% over a very large range of parent temperatures. In stark contrast, mesoscopic elasticity as quantified by $\kappa_g$ varies by more than a factor of 4 over the same $T_p$ range, which is the correct order of magnitude of variation of $\Delta E(T)$ necessary for observing laboratory glass-formers' fragilities \cite{tarjus_2004}. Secondly, the downward bend of $\kappa_g$ with lowering $T_p$, as seen in Fig.~\ref{kappa_g_vs_Tp}b for 2D glasses, could imply that supercooled fragile liquids undergo a \emph{fragile-to-strong transition} on deep supercooling (``strong'' refers to a temperature-independent $\Delta E$), a possibility indeed recently raised by some authors~\cite{COK18,LUCAS2019100034}. As the existence of such a phenomenon would apply a powerful filter on the landscape of theories of the glass transition, its establishment (or refutal) sets an obvious direction for future research.

Another set of observations regarding finite-size effects in plastic activity of stable computer glasses can be linked to our results. Specifically, the characteristic athermal, quasistatic shear-strain $\langle\gamma_1\rangle$ that an as-quenched glassy sample of $N$ particle can undergo before a \emph{first} plastic instability sets in scales as $N^{-\beta}$, with exponents $\beta$ ranging between $[0\!-\!0.70]$ \cite{KLP10b,11HKLP,PhysRevE.92.062302,LW16,OBBRT18,LPRS18,Shang86,kapteijns2019nonlinear}.

Assuming that the distribution $p(\gamma_1)$ inherents its form from the QLM DOS as speculated in Fig.~\ref{fig:DOScartoon} (however, see related discussion in \cite{kapteijns2019nonlinear}), including its $T_p$ dependence, we assert that, as the prefactor $A_g$ of the QLM DOS is reduced upon deeper supercooling, the minimal system size required in order to begin to observe the asymptotic scaling regime $N^\beta$ in the statistics of $\gamma_1$ would increase~\cite{LPRS18}. This minimial size should scale as $\xi_s^\dbar$, where $\xi_s$ was coined the `site length' in \cite{LDB16}, and defined as the characteristic distance between soft QLMs. In \cite{RBL19}, $\xi_s^\dbar$ was estimated to vary by nearly two orders of magnitude for the same glassy samples employed in this work. Since in stable glasses the scaling regime of the QLM DOS ends with the onset of a steep peak, then the mean $\langle\gamma_1\rangle$, given implicitly via \cite{KLP10b}
\begin{equation}
N^{-1} \sim \int_0^{\langle\gamma_1\rangle}p(\gamma_1)d\gamma_1\,,
\end{equation}
is expected to feature a very weak dependence on $N$, if the prefactor $A_g$ and $N$ are small, since the dominant contribution to the integral above will come from the steeply-increasing peak regime of $p(\gamma_1)$, and not from its power-law, scaling regime. We therefore suggest that the small exponents $\beta$ measured in \cite{OBBRT18,Shang86} are a finite-size manifestation of the form of $p(\gamma_1)$, which, in turn, is presumably inherited from ${\cal D}(\omega)$.

\section{Summary}
\label{sec:summary}

In summary, we have shown that the stiffness associated with the normalized linear response of glasses to local force dipoles is a random variable that follows a probability distribution, whose asymptotic form is independent of microscopic details of glass-forming models, and features a weak dependence on spatial dimension. We provided an explanation for the observed statistics of the dipole stiffnesses in terms of a simple lattice model that incorporates both the universal form of the QLM DOS, and QLMs' universal spatial structure. We have subsequently reported calculations of the average dipole response stiffness down to the deeply supercooled regime by availing ourselves to glass-forming models that can be simulated using the Swap Monte Carlo algorithm, and discussed possible conceptual links between our observations and the glass transition problem. By working backwards from these observations and through the insights gained from our lattice model, we proposed a conjecture, schematically summarized in Fig.~\ref{fig:DOScartoon}, on the full form of the density of states of soft, non-phononic excitations in stable (i.e.~low $T_p$) computer glasses, of which only the scaling regime has presently been observed~\cite{LDB16,KBL18}. We further propose that recent observations regarding finite size effects in plastic activity of stable glasses can be explained in terms of our speculations regarding the form of the full QLM DOS. This result concretizes the idea proposed in~\cite{BL18,RBL19}, that dipole responses constitute a useful tool to gain insights into the physics of QLMs in structural glasses.

\acknowledgements
We wish to acknowledge inspiring discussions with G.~D\"uring, G.~Kapteijns and D.~Richard. E.~L.~acknowledges support from the NWO (Vidi grant no.~680-47-554/3259). E.~B.~acknowledges support from the Minerva Foundation with funding from the Federal German Ministry for Education and Research, the Ben May Center for Chemical Theory and Computation, and the Harold Perlman Family.

\appendix

\begin{figure}[!ht]
\includegraphics[width=0.5\textwidth]{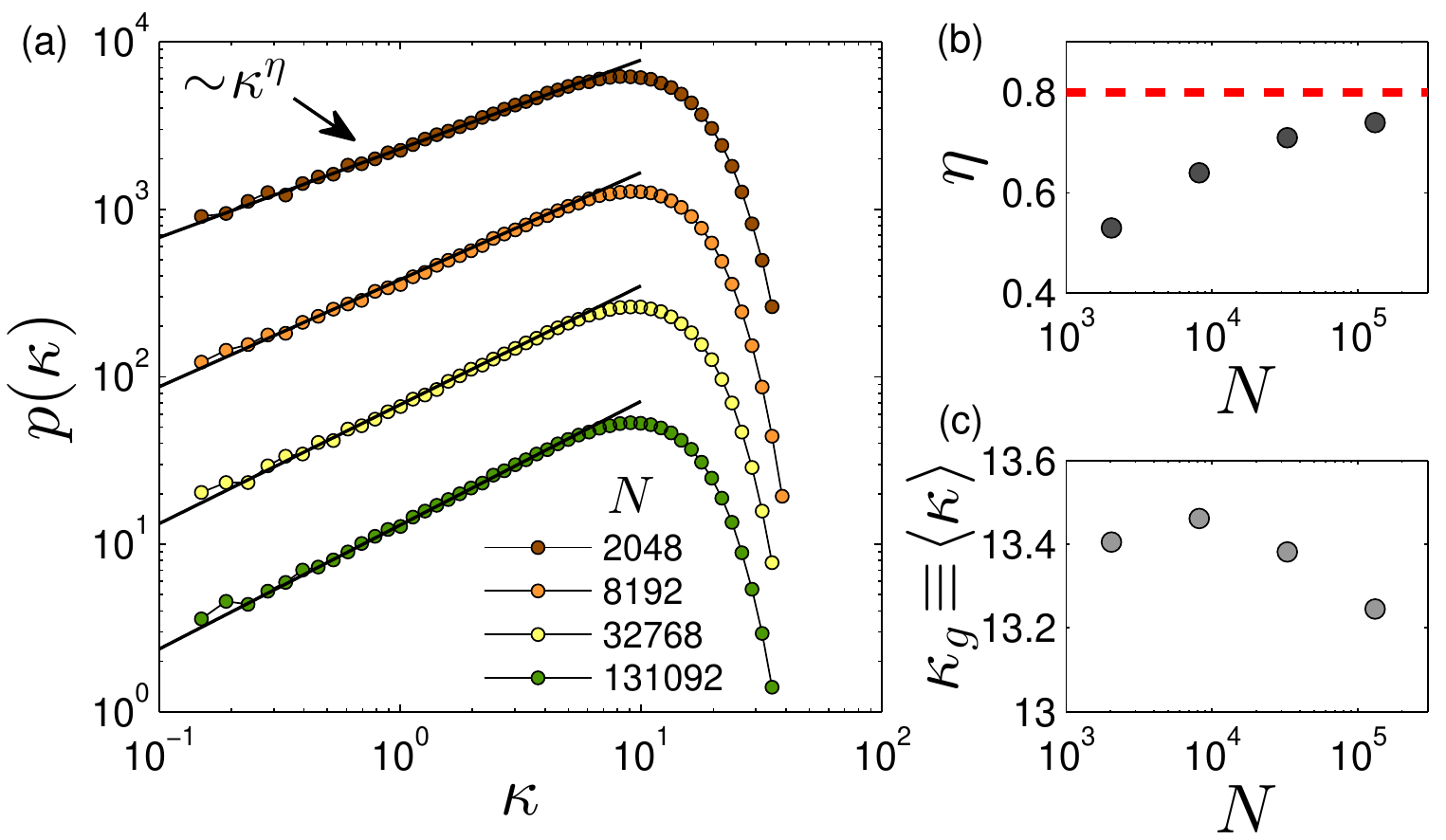}
\caption{Finite size effects in the distributions $p(\kappa)$ of dipole stiffnesses, calculated for $T_p\!=\!1.0$ of the IPL model, and for different system sizes $N$ as detailed in the legend. The exponent appears to slowly converge to the expected $\eta\!=\!0.8$ value reported in the main text, as the system size is increased. We also report how these artifacts (and others, see \cite{LB18}) affect the mean $\kappa_g\equiv\left<\kappa\right>$ itself.}
\label{fig:finitesize}
\end{figure}

\section{Finite size effect on the dipole stiffness' statistics}
\label{app:finitesize}
It was recently reported in~\cite{Lerner19} that the calculation of the QLM DOS  at high $T_p$'s requires extra care be taken to avoid it being contaminated by finite size effects. The reason is that the linear size $\xi_{\rm QLM}$ of localized cores of QLMs grows for glasses quenched from high parent temperatures $T_p$~\cite{LDB16,KBL18,RBL19}. This means that at higher parent temperatures, larger system sizes are needed for the QLMs to be able to fit in the sample and for their universal statistics to be reliably recovered. Similarly, care needs to be taken when considering the statistics of dipole responses and their associated stiffnesses $p(\kappa)$.

We report these finite size effects in Fig.~\ref{fig:finitesize} for the IPL glasses in 3D, quenched instantaneously from high $T_p$, see Sect.~\ref{sec:methods} for details. It can be seen that the scaling exponent is underestimated when the system size is too small, and, upon increasing $N$, approaches the value $\eta\!=\!0.80$ reported for the lattice model in Fig.~\ref{fig:spatial_dependence_glass}a, and for more stable glasses in Figs.~\ref{fig:spatial_dependence_glass}b, \ref{model_independence} and \ref{fig:pdfk_Tp}.


\begin{thebibliography}{53}%
\makeatletter
\providecommand \@ifxundefined [1]{%
 \@ifx{#1\undefined}
}%
\providecommand \@ifnum [1]{%
 \ifnum #1\expandafter \@firstoftwo
 \else \expandafter \@secondoftwo
 \fi
}%
\providecommand \@ifx [1]{%
 \ifx #1\expandafter \@firstoftwo
 \else \expandafter \@secondoftwo
 \fi
}%
\providecommand \natexlab [1]{#1}%
\providecommand \enquote  [1]{``#1''}%
\providecommand \bibnamefont  [1]{#1}%
\providecommand \bibfnamefont [1]{#1}%
\providecommand \citenamefont [1]{#1}%
\providecommand \href@noop [0]{\@secondoftwo}%
\providecommand \href [0]{\begingroup \@sanitize@url \@href}%
\providecommand \@href[1]{\@@startlink{#1}\@@href}%
\providecommand \@@href[1]{\endgroup#1\@@endlink}%
\providecommand \@sanitize@url [0]{\catcode `\\12\catcode `\$12\catcode
  `\&12\catcode `\#12\catcode `\^12\catcode `\_12\catcode `\%12\relax}%
\providecommand \@@startlink[1]{}%
\providecommand \@@endlink[0]{}%
\providecommand \url  [0]{\begingroup\@sanitize@url \@url }%
\providecommand \@url [1]{\endgroup\@href {#1}{\urlprefix }}%
\providecommand \urlprefix  [0]{URL }%
\providecommand \Eprint [0]{\href }%
\providecommand \doibase [0]{http://dx.doi.org/}%
\providecommand \selectlanguage [0]{\@gobble}%
\providecommand \bibinfo  [0]{\@secondoftwo}%
\providecommand \bibfield  [0]{\@secondoftwo}%
\providecommand \translation [1]{[#1]}%
\providecommand \BibitemOpen [0]{}%
\providecommand \bibitemStop [0]{}%
\providecommand \bibitemNoStop [0]{.\EOS\space}%
\providecommand \EOS [0]{\spacefactor3000\relax}%
\providecommand \BibitemShut  [1]{\csname bibitem#1\endcsname}%
\let\auto@bib@innerbib\@empty
\bibitem [{\citenamefont {Landau}\ and\ \citenamefont {Lifshitz}(1959)}]{59LL}%
  \BibitemOpen
  \bibfield  {author} {\bibinfo {author} {\bibfnamefont {L.~D.}\ \bibnamefont
  {Landau}}\ and\ \bibinfo {author} {\bibfnamefont {E.~M.}\ \bibnamefont
  {Lifshitz}},\ }\href@noop {} {\emph {\bibinfo {title} {Course of Theoretical
  Physics Vol 7: Theory of Elasticity}}}\ (\bibinfo  {publisher} {Pergamon
  Press},\ \bibinfo {year} {1959})\BibitemShut {NoStop}%
\bibitem [{\citenamefont {Ashcroft}\ and\ \citenamefont
  {Mermin}(1976)}]{AshcroftMermin}%
  \BibitemOpen
  \bibfield  {author} {\bibinfo {author} {\bibfnamefont {N.~W.}\ \bibnamefont
  {Ashcroft}}\ and\ \bibinfo {author} {\bibfnamefont {N.~D.}\ \bibnamefont
  {Mermin}},\ }\href@noop {} {\emph {\bibinfo {title} {Solid State Physics}}}\
  (\bibinfo  {publisher} {Holt, Rinehart and Winston, New York},\ \bibinfo
  {year} {1976})\BibitemShut {NoStop}%
\bibitem [{\citenamefont {Schober}\ and\ \citenamefont {Laird}(1991)}]{SL91}%
  \BibitemOpen
  \bibfield  {author} {\bibinfo {author} {\bibfnamefont {H.~R.}\ \bibnamefont
  {Schober}}\ and\ \bibinfo {author} {\bibfnamefont {B.~B.}\ \bibnamefont
  {Laird}},\ }\href {\doibase 10.1103/PhysRevB.44.6746} {\bibfield  {journal}
  {\bibinfo  {journal} {Phys. Rev. B}\ }\textbf {\bibinfo {volume} {44}},\
  \bibinfo {pages} {6746} (\bibinfo {year} {1991})}\BibitemShut {NoStop}%
\bibitem [{\citenamefont {Laird}\ and\ \citenamefont {Schober}(1991)}]{LS91}%
  \BibitemOpen
  \bibfield  {author} {\bibinfo {author} {\bibfnamefont {B.~B.}\ \bibnamefont
  {Laird}}\ and\ \bibinfo {author} {\bibfnamefont {H.~R.}\ \bibnamefont
  {Schober}},\ }\href {\doibase 10.1103/PhysRevLett.66.636} {\bibfield
  {journal} {\bibinfo  {journal} {Phys. Rev. Lett.}\ }\textbf {\bibinfo
  {volume} {66}},\ \bibinfo {pages} {636} (\bibinfo {year} {1991})}\BibitemShut
  {NoStop}%
\bibitem [{\citenamefont {Zeller}\ and\ \citenamefont {Pohl}(1971)}]{ZP71}%
  \BibitemOpen
  \bibfield  {author} {\bibinfo {author} {\bibfnamefont {R.~C.}\ \bibnamefont
  {Zeller}}\ and\ \bibinfo {author} {\bibfnamefont {R.~O.}\ \bibnamefont
  {Pohl}},\ }\href {\doibase 10.1103/PhysRevB.4.2029} {\bibfield  {journal}
  {\bibinfo  {journal} {Phys. Rev. B}\ }\textbf {\bibinfo {volume} {4}},\
  \bibinfo {pages} {2029} (\bibinfo {year} {1971})}\BibitemShut {NoStop}%
\bibitem [{\citenamefont {Anderson}\ \emph {et~al.}(1972)\citenamefont
  {Anderson}, \citenamefont {Halperin},\ and\ \citenamefont {Varma}}]{AHV72}%
  \BibitemOpen
  \bibfield  {author} {\bibinfo {author} {\bibfnamefont {P.~W.}\ \bibnamefont
  {Anderson}}, \bibinfo {author} {\bibfnamefont {B.~I.}\ \bibnamefont
  {Halperin}}, \ and\ \bibinfo {author} {\bibfnamefont {C.~M.}\ \bibnamefont
  {Varma}},\ }\href {\doibase 10.1080/14786437208229210} {\bibfield  {journal}
  {\bibinfo  {journal} {Philos. Mag.}\ }\textbf {\bibinfo {volume} {25}},\
  \bibinfo {pages} {1} (\bibinfo {year} {1972})}\BibitemShut {NoStop}%
\bibitem [{\citenamefont {Phillips}(1972)}]{P72}%
  \BibitemOpen
  \bibfield  {author} {\bibinfo {author} {\bibfnamefont {W.}~\bibnamefont
  {Phillips}},\ }\href {\doibase 10.1007/BF00660072} {\bibfield  {journal}
  {\bibinfo  {journal} {J. Low Temp. Phys.}\ }\textbf {\bibinfo {volume} {7}},\
  \bibinfo {pages} {351} (\bibinfo {year} {1972})}\BibitemShut {NoStop}%
\bibitem [{\citenamefont {Manning}\ and\ \citenamefont
  {Liu}(2011)}]{manningliu11}%
  \BibitemOpen
  \bibfield  {author} {\bibinfo {author} {\bibfnamefont {M.~L.}\ \bibnamefont
  {Manning}}\ and\ \bibinfo {author} {\bibfnamefont {A.~J.}\ \bibnamefont
  {Liu}},\ }\href {\doibase 10.1103/PhysRevLett.107.108302} {\bibfield
  {journal} {\bibinfo  {journal} {Phys. Rev. Lett.}\ }\textbf {\bibinfo
  {volume} {107}},\ \bibinfo {pages} {108302} (\bibinfo {year}
  {2011})}\BibitemShut {NoStop}%
\bibitem [{\citenamefont {Rottler}\ \emph {et~al.}(2014)\citenamefont
  {Rottler}, \citenamefont {Schoenholz},\ and\ \citenamefont {Liu}}]{RSSL14}%
  \BibitemOpen
  \bibfield  {author} {\bibinfo {author} {\bibfnamefont {J.}~\bibnamefont
  {Rottler}}, \bibinfo {author} {\bibfnamefont {S.~S.}\ \bibnamefont
  {Schoenholz}}, \ and\ \bibinfo {author} {\bibfnamefont {A.~J.}\ \bibnamefont
  {Liu}},\ }\href {\doibase 10.1103/PhysRevE.89.042304} {\bibfield  {journal}
  {\bibinfo  {journal} {Phys. Rev. E}\ }\textbf {\bibinfo {volume} {89}},\
  \bibinfo {pages} {042304} (\bibinfo {year} {2014})}\BibitemShut {NoStop}%
\bibitem [{\citenamefont {Tanguy}\ \emph {et~al.}(2010)\citenamefont {Tanguy},
  \citenamefont {Mantisi},\ and\ \citenamefont {Tsamados}}]{TMT10}%
  \BibitemOpen
  \bibfield  {author} {\bibinfo {author} {\bibfnamefont {A.}~\bibnamefont
  {Tanguy}}, \bibinfo {author} {\bibfnamefont {B.}~\bibnamefont {Mantisi}}, \
  and\ \bibinfo {author} {\bibfnamefont {M.}~\bibnamefont {Tsamados}},\ }\href
  {http://stacks.iop.org/0295-5075/90/i=1/a=16004} {\bibfield  {journal}
  {\bibinfo  {journal} {Europhys. Lett.}\ }\textbf {\bibinfo {volume} {90}},\
  \bibinfo {pages} {16004} (\bibinfo {year} {2010})}\BibitemShut {NoStop}%
\bibitem [{\citenamefont {Karmakar}\ \emph {et~al.}(2010)\citenamefont
  {Karmakar}, \citenamefont {Lerner},\ and\ \citenamefont
  {Procaccia}}]{KLP10b}%
  \BibitemOpen
  \bibfield  {author} {\bibinfo {author} {\bibfnamefont {S.}~\bibnamefont
  {Karmakar}}, \bibinfo {author} {\bibfnamefont {E.}~\bibnamefont {Lerner}}, \
  and\ \bibinfo {author} {\bibfnamefont {I.}~\bibnamefont {Procaccia}},\ }\href
  {\doibase 10.1103/PhysRevE.82.055103} {\bibfield  {journal} {\bibinfo
  {journal} {Phys. Rev. E}\ }\textbf {\bibinfo {volume} {82}},\ \bibinfo
  {pages} {055103} (\bibinfo {year} {2010})}\BibitemShut {NoStop}%
\bibitem [{\citenamefont {Schober}\ \emph {et~al.}(1993)\citenamefont
  {Schober}, \citenamefont {Oligschleger},\ and\ \citenamefont
  {Laird}}]{schober_1993_numerics}%
  \BibitemOpen
  \bibfield  {author} {\bibinfo {author} {\bibfnamefont {H.}~\bibnamefont
  {Schober}}, \bibinfo {author} {\bibfnamefont {C.}~\bibnamefont
  {Oligschleger}}, \ and\ \bibinfo {author} {\bibfnamefont {B.}~\bibnamefont
  {Laird}},\ }\href {\doibase https://doi.org/10.1016/0022-3093(93)90106-8}
  {\bibfield  {journal} {\bibinfo  {journal} {J. Non-Cryst. Solids}\ }\textbf
  {\bibinfo {volume} {156-158}},\ \bibinfo {pages} {965 } (\bibinfo {year}
  {1993})}\BibitemShut {NoStop}%
\bibitem [{\citenamefont {Brito}\ and\ \citenamefont {Wyart}(2007)}]{BW07}%
  \BibitemOpen
  \bibfield  {author} {\bibinfo {author} {\bibfnamefont {C.}~\bibnamefont
  {Brito}}\ and\ \bibinfo {author} {\bibfnamefont {M.}~\bibnamefont {Wyart}},\
  }\href {http://stacks.iop.org/1742-5468/2007/i=08/a=L08003} {\bibfield
  {journal} {\bibinfo  {journal} {J. Stat. Mech.}\ }\textbf {\bibinfo {volume}
  {2007}},\ \bibinfo {pages} {L08003} (\bibinfo {year} {2007})}\BibitemShut
  {NoStop}%
\bibitem [{\citenamefont {Widmer-Cooper}\ \emph {et~al.}(2008)\citenamefont
  {Widmer-Cooper}, \citenamefont {Perry}, \citenamefont {Harrowell},\ and\
  \citenamefont {Reichman}}]{WPHR08}%
  \BibitemOpen
  \bibfield  {author} {\bibinfo {author} {\bibfnamefont {A.}~\bibnamefont
  {Widmer-Cooper}}, \bibinfo {author} {\bibfnamefont {H.}~\bibnamefont
  {Perry}}, \bibinfo {author} {\bibfnamefont {P.}~\bibnamefont {Harrowell}}, \
  and\ \bibinfo {author} {\bibfnamefont {D.~R.}\ \bibnamefont {Reichman}},\
  }\href {\doibase 10.1038/nphys1025} {\bibfield  {journal} {\bibinfo
  {journal} {Nature Physics}\ }\textbf {\bibinfo {volume} {4}},\ \bibinfo
  {pages} {711} (\bibinfo {year} {2008})}\BibitemShut {NoStop}%
\bibitem [{\citenamefont {Widmer-Cooper}\ \emph {et~al.}(2009)\citenamefont
  {Widmer-Cooper}, \citenamefont {Perry}, \citenamefont {Harrowell},\ and\
  \citenamefont {Reichman}}]{WPHR09}%
  \BibitemOpen
  \bibfield  {author} {\bibinfo {author} {\bibfnamefont {A.}~\bibnamefont
  {Widmer-Cooper}}, \bibinfo {author} {\bibfnamefont {H.}~\bibnamefont
  {Perry}}, \bibinfo {author} {\bibfnamefont {P.}~\bibnamefont {Harrowell}}, \
  and\ \bibinfo {author} {\bibfnamefont {D.~R.}\ \bibnamefont {Reichman}},\
  }\href {\doibase 10.1063/1.3265983} {\bibfield  {journal} {\bibinfo
  {journal} {J. Chem. Phys.}\ }\textbf {\bibinfo {volume} {131}},\ \bibinfo
  {pages} {194508} (\bibinfo {year} {2009})}\BibitemShut {NoStop}%
\bibitem [{\citenamefont {Brito}\ and\ \citenamefont {Wyart}(2009)}]{BW09}%
  \BibitemOpen
  \bibfield  {author} {\bibinfo {author} {\bibfnamefont {C.}~\bibnamefont
  {Brito}}\ and\ \bibinfo {author} {\bibfnamefont {M.}~\bibnamefont {Wyart}},\
  }\href {\doibase 10.1063/1.3157261} {\bibfield  {journal} {\bibinfo
  {journal} {J. Chem. Phys.}\ }\textbf {\bibinfo {volume} {131}},\ \bibinfo
  {pages} {024504} (\bibinfo {year} {2009})}\BibitemShut {NoStop}%
\bibitem [{\citenamefont {Bouchbinder}\ and\ \citenamefont
  {Lerner}(2018)}]{BL18}%
  \BibitemOpen
  \bibfield  {author} {\bibinfo {author} {\bibfnamefont {E.}~\bibnamefont
  {Bouchbinder}}\ and\ \bibinfo {author} {\bibfnamefont {E.}~\bibnamefont
  {Lerner}},\ }\href {http://stacks.iop.org/1367-2630/20/i=7/a=073022}
  {\bibfield  {journal} {\bibinfo  {journal} {New J. Phys.}\ }\textbf {\bibinfo
  {volume} {20}},\ \bibinfo {pages} {073022} (\bibinfo {year}
  {2018})}\BibitemShut {NoStop}%
\bibitem [{\citenamefont {Kapteijns}\ \emph {et~al.}(2019)\citenamefont
  {Kapteijns}, \citenamefont {Richard},\ and\ \citenamefont
  {Lerner}}]{kapteijns2019nonlinear}%
  \BibitemOpen
  \bibfield  {author} {\bibinfo {author} {\bibfnamefont {G.}~\bibnamefont
  {Kapteijns}}, \bibinfo {author} {\bibfnamefont {D.}~\bibnamefont {Richard}},
  \ and\ \bibinfo {author} {\bibfnamefont {E.}~\bibnamefont {Lerner}},\ }\href
  {https://arxiv.org/abs/1912.10930} {\bibfield  {journal} {\bibinfo  {journal}
  {arXiv preprint arXiv:1912.10930}\ } (\bibinfo {year} {2019})}\BibitemShut
  {NoStop}%
\bibitem [{\citenamefont {Lerner}\ \emph {et~al.}(2016)\citenamefont {Lerner},
  \citenamefont {D\"uring},\ and\ \citenamefont {Bouchbinder}}]{LDB16}%
  \BibitemOpen
  \bibfield  {author} {\bibinfo {author} {\bibfnamefont {E.}~\bibnamefont
  {Lerner}}, \bibinfo {author} {\bibfnamefont {G.}~\bibnamefont {D\"uring}}, \
  and\ \bibinfo {author} {\bibfnamefont {E.}~\bibnamefont {Bouchbinder}},\
  }\href {\doibase 10.1103/PhysRevLett.117.035501} {\bibfield  {journal}
  {\bibinfo  {journal} {Phys. Rev. Lett.}\ }\textbf {\bibinfo {volume} {117}},\
  \bibinfo {pages} {035501} (\bibinfo {year} {2016})}\BibitemShut {NoStop}%
\bibitem [{\citenamefont {Kapteijns}\ \emph {et~al.}(2018)\citenamefont
  {Kapteijns}, \citenamefont {Bouchbinder},\ and\ \citenamefont
  {Lerner}}]{KBL18}%
  \BibitemOpen
  \bibfield  {author} {\bibinfo {author} {\bibfnamefont {G.}~\bibnamefont
  {Kapteijns}}, \bibinfo {author} {\bibfnamefont {E.}~\bibnamefont
  {Bouchbinder}}, \ and\ \bibinfo {author} {\bibfnamefont {E.}~\bibnamefont
  {Lerner}},\ }\href {\doibase 10.1103/PhysRevLett.121.055501} {\bibfield
  {journal} {\bibinfo  {journal} {Phys. Rev. Lett.}\ }\textbf {\bibinfo
  {volume} {121}},\ \bibinfo {pages} {055501} (\bibinfo {year}
  {2018})}\BibitemShut {NoStop}%
\bibitem [{\citenamefont {Shimada}\ \emph {et~al.}(2018)\citenamefont
  {Shimada}, \citenamefont {Mizuno},\ and\ \citenamefont {Ikeda}}]{SMI18}%
  \BibitemOpen
  \bibfield  {author} {\bibinfo {author} {\bibfnamefont {M.}~\bibnamefont
  {Shimada}}, \bibinfo {author} {\bibfnamefont {H.}~\bibnamefont {Mizuno}}, \
  and\ \bibinfo {author} {\bibfnamefont {A.}~\bibnamefont {Ikeda}},\ }\href
  {\doibase 10.1103/PhysRevE.97.022609} {\bibfield  {journal} {\bibinfo
  {journal} {Phys. Rev. E}\ }\textbf {\bibinfo {volume} {97}},\ \bibinfo
  {pages} {022609} (\bibinfo {year} {2018})}\BibitemShut {NoStop}%
\bibitem [{\citenamefont {Zylberg}\ \emph {et~al.}(2017)\citenamefont
  {Zylberg}, \citenamefont {Lerner}, \citenamefont {Bar-Sinai},\ and\
  \citenamefont {Bouchbinder}}]{ZLBB17}%
  \BibitemOpen
  \bibfield  {author} {\bibinfo {author} {\bibfnamefont {J.}~\bibnamefont
  {Zylberg}}, \bibinfo {author} {\bibfnamefont {E.}~\bibnamefont {Lerner}},
  \bibinfo {author} {\bibfnamefont {Y.}~\bibnamefont {Bar-Sinai}}, \ and\
  \bibinfo {author} {\bibfnamefont {E.}~\bibnamefont {Bouchbinder}},\ }\href
  {\doibase 10.1073/pnas.1704403114} {\bibfield  {journal} {\bibinfo  {journal}
  {Proc. Natl. Acad. Sci. U.S.A.}\ }\textbf {\bibinfo {volume} {114}},\
  \bibinfo {pages} {7289} (\bibinfo {year} {2017})}\BibitemShut {NoStop}%
\bibitem [{\citenamefont {Lerner}\ and\ \citenamefont
  {Bouchbinder}(2018)}]{LB18}%
  \BibitemOpen
  \bibfield  {author} {\bibinfo {author} {\bibfnamefont {E.}~\bibnamefont
  {Lerner}}\ and\ \bibinfo {author} {\bibfnamefont {E.}~\bibnamefont
  {Bouchbinder}},\ }\href {\doibase 10.1063/1.5024776} {\bibfield  {journal}
  {\bibinfo  {journal} {J. Chem. Phys.}\ }\textbf {\bibinfo {volume} {148}},\
  \bibinfo {pages} {214502} (\bibinfo {year} {2018})}\BibitemShut {NoStop}%
\bibitem [{\citenamefont {{Rainone}}\ \emph {et~al.}(2019)\citenamefont
  {{Rainone}}, \citenamefont {{Bouchbinder}},\ and\ \citenamefont
  {{Lerner}}}]{RBL19}%
  \BibitemOpen
  \bibfield  {author} {\bibinfo {author} {\bibfnamefont {C.}~\bibnamefont
  {{Rainone}}}, \bibinfo {author} {\bibfnamefont {E.}~\bibnamefont
  {{Bouchbinder}}}, \ and\ \bibinfo {author} {\bibfnamefont {E.}~\bibnamefont
  {{Lerner}}},\ }\href {https://arxiv.org/abs/1911.07744} {\bibfield  {journal}
  {\bibinfo  {journal} {Proc. Natl. Acad. Sci. U.S.A., in press. Also
  arXiv:1911.07744}\ } (\bibinfo {year} {2019})}\BibitemShut {NoStop}%
\bibitem [{\citenamefont {O'Hern}\ \emph {et~al.}(2003)\citenamefont {O'Hern},
  \citenamefont {Silbert}, \citenamefont {Liu},\ and\ \citenamefont
  {Nagel}}]{ohern2003}%
  \BibitemOpen
  \bibfield  {author} {\bibinfo {author} {\bibfnamefont {C.~S.}\ \bibnamefont
  {O'Hern}}, \bibinfo {author} {\bibfnamefont {L.~E.}\ \bibnamefont {Silbert}},
  \bibinfo {author} {\bibfnamefont {A.~J.}\ \bibnamefont {Liu}}, \ and\
  \bibinfo {author} {\bibfnamefont {S.~R.}\ \bibnamefont {Nagel}},\ }\href
  {\doibase 10.1103/PhysRevE.68.011306} {\bibfield  {journal} {\bibinfo
  {journal} {Phys. Rev. E}\ }\textbf {\bibinfo {volume} {68}},\ \bibinfo
  {pages} {011306} (\bibinfo {year} {2003})}\BibitemShut {NoStop}%
\bibitem [{\citenamefont {Liu}\ and\ \citenamefont {Nagel}(2010)}]{liu_review}%
  \BibitemOpen
  \bibfield  {author} {\bibinfo {author} {\bibfnamefont {A.~J.}\ \bibnamefont
  {Liu}}\ and\ \bibinfo {author} {\bibfnamefont {S.~R.}\ \bibnamefont
  {Nagel}},\ }\href {\doibase 10.1146/annurev-conmatphys-070909-104045}
  {\bibfield  {journal} {\bibinfo  {journal} {Annu. Rev. Condens. Matter
  Phys.}\ }\textbf {\bibinfo {volume} {1}},\ \bibinfo {pages} {347} (\bibinfo
  {year} {2010})}\BibitemShut {NoStop}%
\bibitem [{\citenamefont {van Hecke}(2010)}]{van_hecke_review}%
  \BibitemOpen
  \bibfield  {author} {\bibinfo {author} {\bibfnamefont {M.}~\bibnamefont {van
  Hecke}},\ }\href {http://stacks.iop.org/0953-8984/22/i=3/a=033101} {\bibfield
   {journal} {\bibinfo  {journal} {J. Phys.: Condens. Matter}\ }\textbf
  {\bibinfo {volume} {22}},\ \bibinfo {pages} {033101} (\bibinfo {year}
  {2010})}\BibitemShut {NoStop}%
\bibitem [{\citenamefont {Yan}\ \emph {et~al.}(2016)\citenamefont {Yan},
  \citenamefont {DeGiuli},\ and\ \citenamefont
  {Wyart}}]{new_variational_argument_epl_2016}%
  \BibitemOpen
  \bibfield  {author} {\bibinfo {author} {\bibfnamefont {L.}~\bibnamefont
  {Yan}}, \bibinfo {author} {\bibfnamefont {E.}~\bibnamefont {DeGiuli}}, \ and\
  \bibinfo {author} {\bibfnamefont {M.}~\bibnamefont {Wyart}},\ }\href
  {http://stacks.iop.org/0295-5075/114/i=2/a=26003} {\bibfield  {journal}
  {\bibinfo  {journal} {Europhys. Lett.}\ }\textbf {\bibinfo {volume} {114}},\
  \bibinfo {pages} {26003} (\bibinfo {year} {2016})}\BibitemShut {NoStop}%
\bibitem [{\citenamefont {Ninarello}\ \emph {et~al.}(2017)\citenamefont
  {Ninarello}, \citenamefont {Berthier},\ and\ \citenamefont
  {Coslovich}}]{NBC17}%
  \BibitemOpen
  \bibfield  {author} {\bibinfo {author} {\bibfnamefont {A.}~\bibnamefont
  {Ninarello}}, \bibinfo {author} {\bibfnamefont {L.}~\bibnamefont {Berthier}},
  \ and\ \bibinfo {author} {\bibfnamefont {D.}~\bibnamefont {Coslovich}},\
  }\href {\doibase 10.1103/PhysRevX.7.021039} {\bibfield  {journal} {\bibinfo
  {journal} {Phys. Rev. X}\ }\textbf {\bibinfo {volume} {7}},\ \bibinfo {pages}
  {021039} (\bibinfo {year} {2017})}\BibitemShut {NoStop}%
\bibitem [{\citenamefont {Grigera}\ and\ \citenamefont
  {Parisi}(2001)}]{GrigeraParisi01}%
  \BibitemOpen
  \bibfield  {author} {\bibinfo {author} {\bibfnamefont {T.~S.}\ \bibnamefont
  {Grigera}}\ and\ \bibinfo {author} {\bibfnamefont {G.}~\bibnamefont
  {Parisi}},\ }\href {\doibase 10.1103/PhysRevE.63.045102} {\bibfield
  {journal} {\bibinfo  {journal} {Phys. Rev. E}\ }\textbf {\bibinfo {volume}
  {63}},\ \bibinfo {pages} {045102} (\bibinfo {year} {2001})}\BibitemShut
  {NoStop}%
\bibitem [{\citenamefont {Guti\'errez}\ \emph {et~al.}(2015)\citenamefont
  {Guti\'errez}, \citenamefont {Karmakar}, \citenamefont {Pollack},\ and\
  \citenamefont {Procaccia}}]{GKPP15}%
  \BibitemOpen
  \bibfield  {author} {\bibinfo {author} {\bibfnamefont {R.}~\bibnamefont
  {Guti\'errez}}, \bibinfo {author} {\bibfnamefont {S.}~\bibnamefont
  {Karmakar}}, \bibinfo {author} {\bibfnamefont {Y.~G.}\ \bibnamefont
  {Pollack}}, \ and\ \bibinfo {author} {\bibfnamefont {I.}~\bibnamefont
  {Procaccia}},\ }\href {http://stacks.iop.org/0295-5075/111/i=5/a=56009}
  {\bibfield  {journal} {\bibinfo  {journal} {Europhys. Lett.}\ }\textbf
  {\bibinfo {volume} {111}},\ \bibinfo {pages} {56009} (\bibinfo {year}
  {2015})}\BibitemShut {NoStop}%
\bibitem [{\citenamefont {Berthier}\ \emph
  {et~al.}(2019{\natexlab{a}})\citenamefont {Berthier}, \citenamefont
  {Charbonneau}, \citenamefont {Ninarello}, \citenamefont {Ozawa},\ and\
  \citenamefont {Yaida}}]{BCN18}%
  \BibitemOpen
  \bibfield  {author} {\bibinfo {author} {\bibfnamefont {L.}~\bibnamefont
  {Berthier}}, \bibinfo {author} {\bibfnamefont {P.}~\bibnamefont
  {Charbonneau}}, \bibinfo {author} {\bibfnamefont {A.}~\bibnamefont
  {Ninarello}}, \bibinfo {author} {\bibfnamefont {M.}~\bibnamefont {Ozawa}}, \
  and\ \bibinfo {author} {\bibfnamefont {S.}~\bibnamefont {Yaida}},\ }\href
  {\doibase 10.1038/s41467-019-09512-3} {\bibfield  {journal} {\bibinfo
  {journal} {Nat. Commun.}\ }\textbf {\bibinfo {volume} {10}},\ \bibinfo
  {pages} {1508} (\bibinfo {year} {2019}{\natexlab{a}})}\BibitemShut {NoStop}%
\bibitem [{\citenamefont {Berthier}\ \emph
  {et~al.}(2019{\natexlab{b}})\citenamefont {Berthier}, \citenamefont
  {Charbonneau},\ and\ \citenamefont {Kundu}}]{BCK19}%
  \BibitemOpen
  \bibfield  {author} {\bibinfo {author} {\bibfnamefont {L.}~\bibnamefont
  {Berthier}}, \bibinfo {author} {\bibfnamefont {P.}~\bibnamefont
  {Charbonneau}}, \ and\ \bibinfo {author} {\bibfnamefont {J.}~\bibnamefont
  {Kundu}},\ }\href {\doibase 10.1103/PhysRevE.99.031301} {\bibfield  {journal}
  {\bibinfo  {journal} {Phys. Rev. E}\ }\textbf {\bibinfo {volume} {99}},\
  \bibinfo {pages} {031301} (\bibinfo {year} {2019}{\natexlab{b}})}\BibitemShut
  {NoStop}%
\bibitem [{\citenamefont {Lerner}(2019)}]{boring_paper}%
  \BibitemOpen
  \bibfield  {author} {\bibinfo {author} {\bibfnamefont {E.}~\bibnamefont
  {Lerner}},\ }\href {\doibase
  https://doi.org/10.1016/j.jnoncrysol.2019.119570} {\bibfield  {journal}
  {\bibinfo  {journal} {J. Non-Cryst. Solids}\ }\textbf {\bibinfo {volume}
  {522}},\ \bibinfo {pages} {119570} (\bibinfo {year} {2019})}\BibitemShut
  {NoStop}%
\bibitem [{\citenamefont {Kob}\ and\ \citenamefont
  {Andersen}(1995)}]{KobAndersen95}%
  \BibitemOpen
  \bibfield  {author} {\bibinfo {author} {\bibfnamefont {W.}~\bibnamefont
  {Kob}}\ and\ \bibinfo {author} {\bibfnamefont {H.~C.}\ \bibnamefont
  {Andersen}},\ }\href {\doibase 10.1103/PhysRevE.51.4626} {\bibfield
  {journal} {\bibinfo  {journal} {Phys. Rev. E}\ }\textbf {\bibinfo {volume}
  {51}},\ \bibinfo {pages} {4626} (\bibinfo {year} {1995})}\BibitemShut
  {NoStop}%
\bibitem [{\citenamefont {Bouchaud}\ and\ \citenamefont
  {Georges}(1990)}]{BOUCHAUD1990127}%
  \BibitemOpen
  \bibfield  {author} {\bibinfo {author} {\bibfnamefont {J.-P.}\ \bibnamefont
  {Bouchaud}}\ and\ \bibinfo {author} {\bibfnamefont {A.}~\bibnamefont
  {Georges}},\ }\href {\doibase https://doi.org/10.1016/0370-1573(90)90099-N}
  {\bibfield  {journal} {\bibinfo  {journal} {Phys. Rep.}\ }\textbf {\bibinfo
  {volume} {195}},\ \bibinfo {pages} {127 } (\bibinfo {year}
  {1990})}\BibitemShut {NoStop}%
\bibitem [{\citenamefont {Florescu}\ and\ \citenamefont
  {Tudor}(2013)}]{ProbHandbook}%
  \BibitemOpen
  \bibfield  {author} {\bibinfo {author} {\bibfnamefont {I.}~\bibnamefont
  {Florescu}}\ and\ \bibinfo {author} {\bibfnamefont {C.~A.}\ \bibnamefont
  {Tudor}},\ }\href@noop {} {\emph {\bibinfo {title} {Handbook of
  probability}}}\ (\bibinfo  {publisher} {John Wiley \& Sons},\ \bibinfo {year}
  {2013})\BibitemShut {NoStop}%
\bibitem [{\citenamefont {Wang}\ \emph {et~al.}(2019)\citenamefont {Wang},
  \citenamefont {Ninarello}, \citenamefont {Guan}, \citenamefont {Berthier},
  \citenamefont {Szamel},\ and\ \citenamefont {Flenner}}]{WNGBSF18}%
  \BibitemOpen
  \bibfield  {author} {\bibinfo {author} {\bibfnamefont {L.}~\bibnamefont
  {Wang}}, \bibinfo {author} {\bibfnamefont {A.}~\bibnamefont {Ninarello}},
  \bibinfo {author} {\bibfnamefont {P.}~\bibnamefont {Guan}}, \bibinfo {author}
  {\bibfnamefont {L.}~\bibnamefont {Berthier}}, \bibinfo {author}
  {\bibfnamefont {G.}~\bibnamefont {Szamel}}, \ and\ \bibinfo {author}
  {\bibfnamefont {E.}~\bibnamefont {Flenner}},\ }\href {\doibase
  10.1038/s41467-018-07978-1} {\bibfield  {journal} {\bibinfo  {journal} {Nat.
  Commun.}\ }\textbf {\bibinfo {volume} {10}},\ \bibinfo {pages} {26} (\bibinfo
  {year} {2019})}\BibitemShut {NoStop}%
\bibitem [{foo()}]{footnote}%
  \BibitemOpen
  \href@noop {} {\bibfield  {journal} {\bibinfo  {journal} {In
  \cite{KBL18,RBL19} it was shown that for a very particular set of
  circumstances (large $\dbar$, small $N$, high $T_p$), the breakdown of the
  scaling regime of the universal ${\cal D}(\omega)\!\sim\!\omega^4$ can be
  seen in the harmonic density of states}\ } (\bibinfo {year} {.})}\BibitemShut
  {NoStop}%
\bibitem [{\citenamefont {Lema{\^i}tre}\ and\ \citenamefont
  {Maloney}(2006)}]{LM06}%
  \BibitemOpen
  \bibfield  {author} {\bibinfo {author} {\bibfnamefont {A.}~\bibnamefont
  {Lema{\^i}tre}}\ and\ \bibinfo {author} {\bibfnamefont {C.}~\bibnamefont
  {Maloney}},\ }\href {\doibase 10.1007/s10955-005-9015-5} {\bibfield
  {journal} {\bibinfo  {journal} {J. Stat. Phys.}\ }\textbf {\bibinfo {volume}
  {123}},\ \bibinfo {pages} {415} (\bibinfo {year} {2006})}\BibitemShut
  {NoStop}%
\bibitem [{\citenamefont {Cavagna}(2009)}]{CavagnaLiq}%
  \BibitemOpen
  \bibfield  {author} {\bibinfo {author} {\bibfnamefont {A.}~\bibnamefont
  {Cavagna}},\ }\href {\doibase
  http://dx.doi.org/10.1016/j.physrep.2009.03.003} {\bibfield  {journal}
  {\bibinfo  {journal} {Phys. Rep.}\ }\textbf {\bibinfo {volume} {476}},\
  \bibinfo {pages} {51 } (\bibinfo {year} {2009})}\BibitemShut {NoStop}%
\bibitem [{\citenamefont {Tarjus}\ \emph {et~al.}(2004)\citenamefont {Tarjus},
  \citenamefont {Kivelson}, \citenamefont {Mossa},\ and\ \citenamefont
  {Alba-Simionesco}}]{tarjus_2004}%
  \BibitemOpen
  \bibfield  {author} {\bibinfo {author} {\bibfnamefont {G.}~\bibnamefont
  {Tarjus}}, \bibinfo {author} {\bibfnamefont {D.}~\bibnamefont {Kivelson}},
  \bibinfo {author} {\bibfnamefont {S.}~\bibnamefont {Mossa}}, \ and\ \bibinfo
  {author} {\bibfnamefont {C.}~\bibnamefont {Alba-Simionesco}},\ }\href
  {\doibase 10.1063/1.1649732} {\bibfield  {journal} {\bibinfo  {journal} {J.
  Chem. Phys.}\ }\textbf {\bibinfo {volume} {120}},\ \bibinfo {pages} {6135}
  (\bibinfo {year} {2004})}\BibitemShut {NoStop}%
\bibitem [{\citenamefont {Dyre}(2006)}]{Dy06}%
  \BibitemOpen
  \bibfield  {author} {\bibinfo {author} {\bibfnamefont {J.~C.}\ \bibnamefont
  {Dyre}},\ }\href {\doibase 10.1103/RevModPhys.78.953} {\bibfield  {journal}
  {\bibinfo  {journal} {Rev. Mod. Phys.}\ }\textbf {\bibinfo {volume} {78}},\
  \bibinfo {pages} {953} (\bibinfo {year} {2006})}\BibitemShut {NoStop}%
\bibitem [{\citenamefont {Dyre}\ \emph {et~al.}(1996)\citenamefont {Dyre},
  \citenamefont {Olsen},\ and\ \citenamefont {Christensen}}]{PhysRevB.53.2171}%
  \BibitemOpen
  \bibfield  {author} {\bibinfo {author} {\bibfnamefont {J.~C.}\ \bibnamefont
  {Dyre}}, \bibinfo {author} {\bibfnamefont {N.~B.}\ \bibnamefont {Olsen}}, \
  and\ \bibinfo {author} {\bibfnamefont {T.}~\bibnamefont {Christensen}},\
  }\href {\doibase 10.1103/PhysRevB.53.2171} {\bibfield  {journal} {\bibinfo
  {journal} {Phys. Rev. B}\ }\textbf {\bibinfo {volume} {53}},\ \bibinfo
  {pages} {2171} (\bibinfo {year} {1996})}\BibitemShut {NoStop}%
\bibitem [{\citenamefont {Coslovich}\ \emph {et~al.}(2018)\citenamefont
  {Coslovich}, \citenamefont {Ozawa},\ and\ \citenamefont {Kob}}]{COK18}%
  \BibitemOpen
  \bibfield  {author} {\bibinfo {author} {\bibfnamefont {D.}~\bibnamefont
  {Coslovich}}, \bibinfo {author} {\bibfnamefont {M.}~\bibnamefont {Ozawa}}, \
  and\ \bibinfo {author} {\bibfnamefont {W.}~\bibnamefont {Kob}},\ }\href
  {\doibase 10.1140/epje/i2018-11671-2} {\bibfield  {journal} {\bibinfo
  {journal} {Eur. Phys. J. E}\ }\textbf {\bibinfo {volume} {41}},\ \bibinfo
  {pages} {62} (\bibinfo {year} {2018})}\BibitemShut {NoStop}%
\bibitem [{\citenamefont {Lucas}(2019)}]{LUCAS2019100034}%
  \BibitemOpen
  \bibfield  {author} {\bibinfo {author} {\bibfnamefont {P.}~\bibnamefont
  {Lucas}},\ }\href {\doibase https://doi.org/10.1016/j.nocx.2019.100034}
  {\bibfield  {journal} {\bibinfo  {journal} {J. Non-Cryst. Solids}\ }\textbf
  {\bibinfo {volume} {4}},\ \bibinfo {pages} {100034} (\bibinfo {year}
  {2019})}\BibitemShut {NoStop}%
\bibitem [{\citenamefont {Hentschel}\ \emph {et~al.}(2011)\citenamefont
  {Hentschel}, \citenamefont {Karmakar}, \citenamefont {Lerner},\ and\
  \citenamefont {Procaccia}}]{11HKLP}%
  \BibitemOpen
  \bibfield  {author} {\bibinfo {author} {\bibfnamefont {H.~G.~E.}\
  \bibnamefont {Hentschel}}, \bibinfo {author} {\bibfnamefont {S.}~\bibnamefont
  {Karmakar}}, \bibinfo {author} {\bibfnamefont {E.}~\bibnamefont {Lerner}}, \
  and\ \bibinfo {author} {\bibfnamefont {I.}~\bibnamefont {Procaccia}},\ }\href
  {\doibase 10.1103/PhysRevE.83.061101} {\bibfield  {journal} {\bibinfo
  {journal} {Phys. Rev. E}\ }\textbf {\bibinfo {volume} {83}},\ \bibinfo
  {pages} {061101} (\bibinfo {year} {2011})}\BibitemShut {NoStop}%
\bibitem [{\citenamefont {Hentschel}\ \emph {et~al.}(2015)\citenamefont
  {Hentschel}, \citenamefont {Jaiswal}, \citenamefont {Procaccia},\ and\
  \citenamefont {Sastry}}]{PhysRevE.92.062302}%
  \BibitemOpen
  \bibfield  {author} {\bibinfo {author} {\bibfnamefont {H.~G.~E.}\
  \bibnamefont {Hentschel}}, \bibinfo {author} {\bibfnamefont {P.~K.}\
  \bibnamefont {Jaiswal}}, \bibinfo {author} {\bibfnamefont {I.}~\bibnamefont
  {Procaccia}}, \ and\ \bibinfo {author} {\bibfnamefont {S.}~\bibnamefont
  {Sastry}},\ }\href {\doibase 10.1103/PhysRevE.92.062302} {\bibfield
  {journal} {\bibinfo  {journal} {Phys. Rev. E}\ }\textbf {\bibinfo {volume}
  {92}},\ \bibinfo {pages} {062302} (\bibinfo {year} {2015})}\BibitemShut
  {NoStop}%
\bibitem [{\citenamefont {Lin}\ and\ \citenamefont {Wyart}(2016)}]{LW16}%
  \BibitemOpen
  \bibfield  {author} {\bibinfo {author} {\bibfnamefont {J.}~\bibnamefont
  {Lin}}\ and\ \bibinfo {author} {\bibfnamefont {M.}~\bibnamefont {Wyart}},\
  }\href {\doibase 10.1103/PhysRevX.6.011005} {\bibfield  {journal} {\bibinfo
  {journal} {Phys. Rev. X}\ }\textbf {\bibinfo {volume} {6}},\ \bibinfo {pages}
  {011005} (\bibinfo {year} {2016})}\BibitemShut {NoStop}%
\bibitem [{\citenamefont {Ozawa}\ \emph {et~al.}(2018)\citenamefont {Ozawa},
  \citenamefont {Berthier}, \citenamefont {Biroli}, \citenamefont {Rosso},\
  and\ \citenamefont {Tarjus}}]{OBBRT18}%
  \BibitemOpen
  \bibfield  {author} {\bibinfo {author} {\bibfnamefont {M.}~\bibnamefont
  {Ozawa}}, \bibinfo {author} {\bibfnamefont {L.}~\bibnamefont {Berthier}},
  \bibinfo {author} {\bibfnamefont {G.}~\bibnamefont {Biroli}}, \bibinfo
  {author} {\bibfnamefont {A.}~\bibnamefont {Rosso}}, \ and\ \bibinfo {author}
  {\bibfnamefont {G.}~\bibnamefont {Tarjus}},\ }\href {\doibase
  10.1073/pnas.1806156115} {\bibfield  {journal} {\bibinfo  {journal} {Proc.
  Natl. Acad. Sci. U.S.A}\ }\textbf {\bibinfo {volume} {115}},\ \bibinfo
  {pages} {6656} (\bibinfo {year} {2018})}\BibitemShut {NoStop}%
\bibitem [{\citenamefont {Lerner}\ \emph {et~al.}(2018)\citenamefont {Lerner},
  \citenamefont {Procaccia}, \citenamefont {Rainone},\ and\ \citenamefont
  {Singh}}]{LPRS18}%
  \BibitemOpen
  \bibfield  {author} {\bibinfo {author} {\bibfnamefont {E.}~\bibnamefont
  {Lerner}}, \bibinfo {author} {\bibfnamefont {I.}~\bibnamefont {Procaccia}},
  \bibinfo {author} {\bibfnamefont {C.}~\bibnamefont {Rainone}}, \ and\
  \bibinfo {author} {\bibfnamefont {M.}~\bibnamefont {Singh}},\ }\href
  {\doibase 10.1103/PhysRevE.98.063001} {\bibfield  {journal} {\bibinfo
  {journal} {Phys. Rev. E}\ }\textbf {\bibinfo {volume} {98}},\ \bibinfo
  {pages} {063001} (\bibinfo {year} {2018})}\BibitemShut {NoStop}%
\bibitem [{\citenamefont {Shang}\ \emph {et~al.}(2020)\citenamefont {Shang},
  \citenamefont {Guan},\ and\ \citenamefont {Barrat}}]{Shang86}%
  \BibitemOpen
  \bibfield  {author} {\bibinfo {author} {\bibfnamefont {B.}~\bibnamefont
  {Shang}}, \bibinfo {author} {\bibfnamefont {P.}~\bibnamefont {Guan}}, \ and\
  \bibinfo {author} {\bibfnamefont {J.-L.}\ \bibnamefont {Barrat}},\ }\href
  {\doibase 10.1073/pnas.1915070117} {\bibfield  {journal} {\bibinfo  {journal}
  {Proc. Natl. Acad. Sci. U.S.A.}\ }\textbf {\bibinfo {volume} {117}},\
  \bibinfo {pages} {86} (\bibinfo {year} {2020})}\BibitemShut {NoStop}%
\bibitem [{\citenamefont {{Lerner}}(2019)}]{Lerner19}%
  \BibitemOpen
  \bibfield  {author} {\bibinfo {author} {\bibfnamefont {E.}~\bibnamefont
  {{Lerner}}},\ }\href {https://arxiv.org/abs/1911.07744} {\bibfield  {journal}
  {\bibinfo  {journal} {arXiv e-prints arXiv:1911.07744}\ } (\bibinfo {year}
  {2019})}\BibitemShut {NoStop}%
\end{thebibliography}
\end{document}